\documentclass[article,11pt]{emulateapj}
\usepackage{natbib}
\usepackage{graphicx}
\usepackage{amsmath}
\usepackage{amssymb}
\def\ao{Appl. Optics }

\def\chisq{$\chi^2$ }
\def\chisqx{$\chi^2$}
\def\chf{CH$_4$ }
\def\chfx{CH$_4$}

\def\cthtx{C$_2$H$_2$}
\def\cbm{cr\`eme br\^ul\'ee }
\def\Cbm{Cr\`eme br\^ul\'ee }

\def\Wm2{W/m$^2$}
\def\Wpm2sr{Wm$^{-2}sr^{-1}$}
\def\deg{$^\circ$ }
\def\degx{$^\circ$}

\def\jgr{J. Geophys. Res. }

\def\mum{$\mu$m }
\def\mumx{$\mu$m}

\def\nht{NH$_3$ }
\def\nhtx{NH$_3$}
\def\nhfsh{NH$_4$SH }
\def\nhfshx{NH$_4$SH}

\begin{document}

\title{A possibly universal red chromophore for modeling color variations on Jupiter}
\author{L.~A. Sromovsky$^1$, K.~H. Baines$^1$, P.~M. Fry$^1$, R.~W. Carlson$^2$}
\affil{$^1$Space Science and Engineering Center, University of Wisconsin-Madison,
1225 West Dayton Street, Madison, WI 53706, USA}
\affil{$^2$Jet Propulsion Laboratory, 4800 Oak Grove Drive, Pasadena, CA 91109, USA}
%
\slugcomment{Journal reference: Icarus 291 (2017) 232-244}

\begin{abstract}
A new laboratory-generated chemical compound made from
photodissociated ammonia (\nhtx) molecules reacting with acetylene
(\cthtx) was suggested as a possible coloring agent for Jupiter's
Great Red Spot (GRS) by Carlson et al. (2016, Icarus 274, 106-115).
Baines et al. (2016, Icarus, submitted) showed that the GRS spectrum
measured by the visual channels of the Cassini VIMS instrument in 2000
could be accurately fit by a cloud model in which the chromophore
appeared as a physically thin layer of small particles immediately
above the main cloud layer of the GRS.  Here we show that the same
chromophore and same layer location can also provide close matches to
the short wave spectra of many other cloud features on Jupiter,
suggesting this material may be a nearly universal chromophore that
could explain the various degrees of red coloration on Jupiter. This
is a robust conclusion, even for 12\% changes in VIMS calibration and
large uncertainties in the refractive index of the main cloud layer
due to uncertain fractions of \nhfsh and \nht in its cloud particles.
The chromophore layer can account for color variations among north and
south equatorial belts, equatorial zone, and the Great Red Spot, by
varying particle size from 0.12 \mum to 0.29 \mum and 1-\mum optical depth
from 0.06 to 0.76.  The total mass of the chromophore layer is much
less variable, ranging from 18 to 30 $\mu$g/cm$^2$, except in the
equatorial zone, where it is only 10-13 $\mu$g/cm$^2$. We also found a
depression of the ammonia volume mixing ratio in the two belt regions,
which averaged 0.4-0.5 $\times 10^{-4}$ immediately below the ammonia
condensation level, while the other regions averaged twice that value.

\end{abstract}
\maketitle
\shortauthors{Sromovsky et al.} 
\shorttitle{A universal chromophore for Jupiter.}
\newpage

\section{Introduction}

\begin{figure}\centering
\hspace{-0.15in}\includegraphics[width=3.45in]{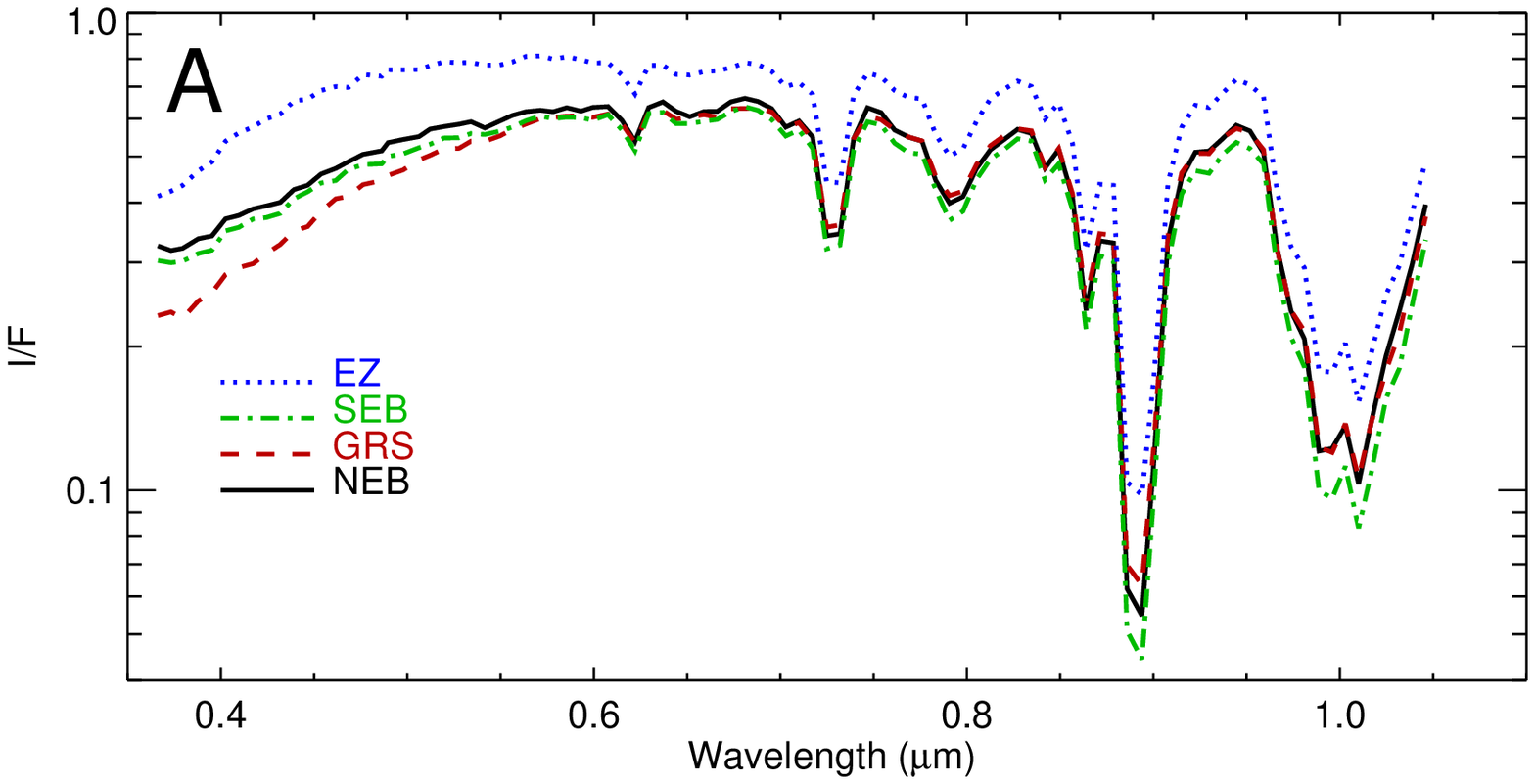}
\includegraphics[width=3.35in]{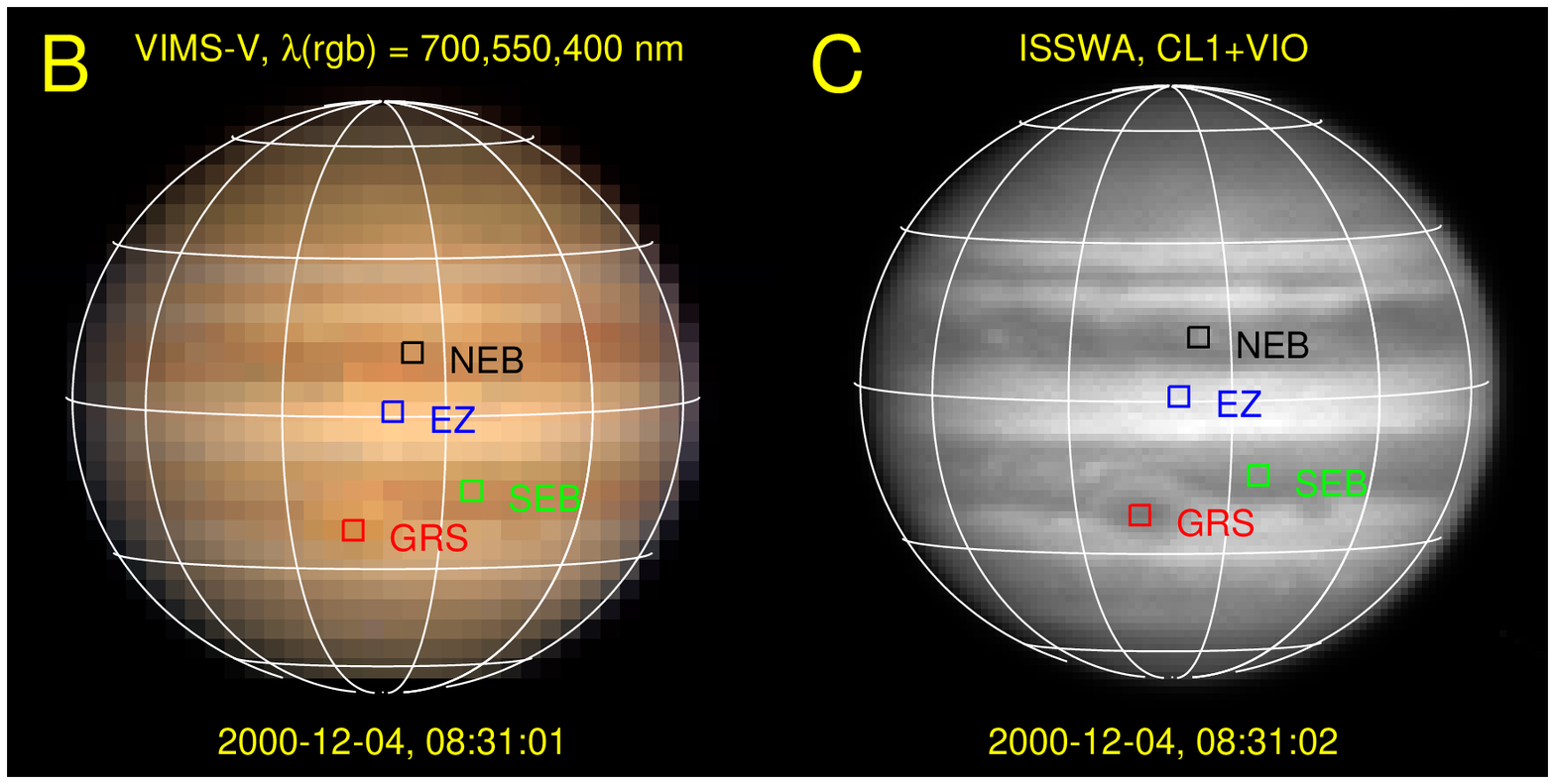}
\caption{A: VIMS spectra from locations indicated in the VIMS image
  composite (B) and the ISS Violet-filtered image (C).  The VIMS color
  composite image of Jupiter on 4 December 2000 used wavelength
  assignments given in the legend. The corresponding ISS image is
  nearly simultaneous with the VIMS observation. The grids are spaced
  30\deg in planetocentric latitude and longitude. Overplotted squares
  indicate locations of spectral samples used to characterize the GRS
  (red, dashed), SEB (green, dot-dash), EZ (blue, dotted) and NEB
  (black, solid).}\label{Fig:lowphase}
\end{figure}

It has long been known that the condensable molecules near the visible
cloud level in Jupiter's atmosphere, including ammonia and ammonium
hydrosulfide, are colorless at visible wavelengths, while Jupiter's
cloud features have an overall red coloration to varying degrees, as
evident from the spectral samples shown in Fig.\ \ref{Fig:lowphase}.
Jupiter's clouds presumably contain some unknown compound that absorbs
blue light preferentially, with the Great Red Spot being a region of
enhanced red coloration.  A number of suggestions have been made over
the years to explain the color of the GRS, including molecules
involving nitrogen, sulfur, phosphorous, and various compounds
generated by irradiation, and complex organics of unknown composition
such as tholins, as summarized by \cite{West1986} and further reviewed
by \cite{West2009satbook}.  Recent arguments
have been advanced for irradiated \nhfsh by \cite{Loeffler2016}.
Until recently no accurate match to the GRS spectrum had ever been
demonstrated.  Judging on the basis of spectral fit quality, the most
promising material suggested as the GRS coloring agent is a
laboratory-generated chemical compound made from photodissociated
ammonia (\nhtx) molecules reacting with acetylene (\cthtx), described
by \cite{Carlson2016}.  \cite{Baines2016Icarus} showed that the GRS
spectrum measured by the visual channels of the Cassini VIMS
instrument in 2000 could be accurately fit by a cloud model in which
the chromophore appeared as a physically thin layer of small chromophore particles
immediately above the main cloud layer of the GRS, which they referred
to as the \cbm model because of the dessert's analogous vertical
structure.  They also considered other models in which the chromophore
appeared in a vertically detached stratospheric haze, which did not fit
as well, or as a coating on the particles of the main cloud layer, for which
the fit was significantly worse.  

Here we use the same VIMS data set, but extend the analysis to other
cloud features and consider more varied vertical structures, showing
that models using the same chromophore can fit much of the color variation that
is normally seen over Jupiter's disk.

\section{Observations}

\subsection{VIMS instrumental characteristics}

 The VIMS instrument \citep{Brown2004} provides two overlapping spectral channels
 covering the ranges from 0.3 to 1.05 $\mu$m (VIS) and from 0.86 to 5
 $\mu$m (IR) with an effective pixel size of 0.5 milliradians on a
 side and a near-IR spectral resolution of approximately 15 nm
 (sampled at intervals of approximately 16 nm).  The IR channel
uses a linear detector to record a spectrum for a single spatial
pixel, so that an image must be acquired by scanning the FOV across
the target.  The visual observations use a CCD matrix detector which records 
both spectral and spatial information simultaneously.  The image of the target is focused on
an entrance slit which is dispersed by a grating and focused on
a two-dimension CCD detector array, recording spatial information
along the slit direction but spectral information in the dispersion
direction. In the normal mode of operation on-chip summing is used
to achieve a spectral resolution of 7.3 nm and a spatial resolution
of 0.5 milliradians.

\subsection{Cassini VIMS observations of Jupiter.}

\begin{figure}\centering
\hspace{-0.15in}\includegraphics[width=3.45in]{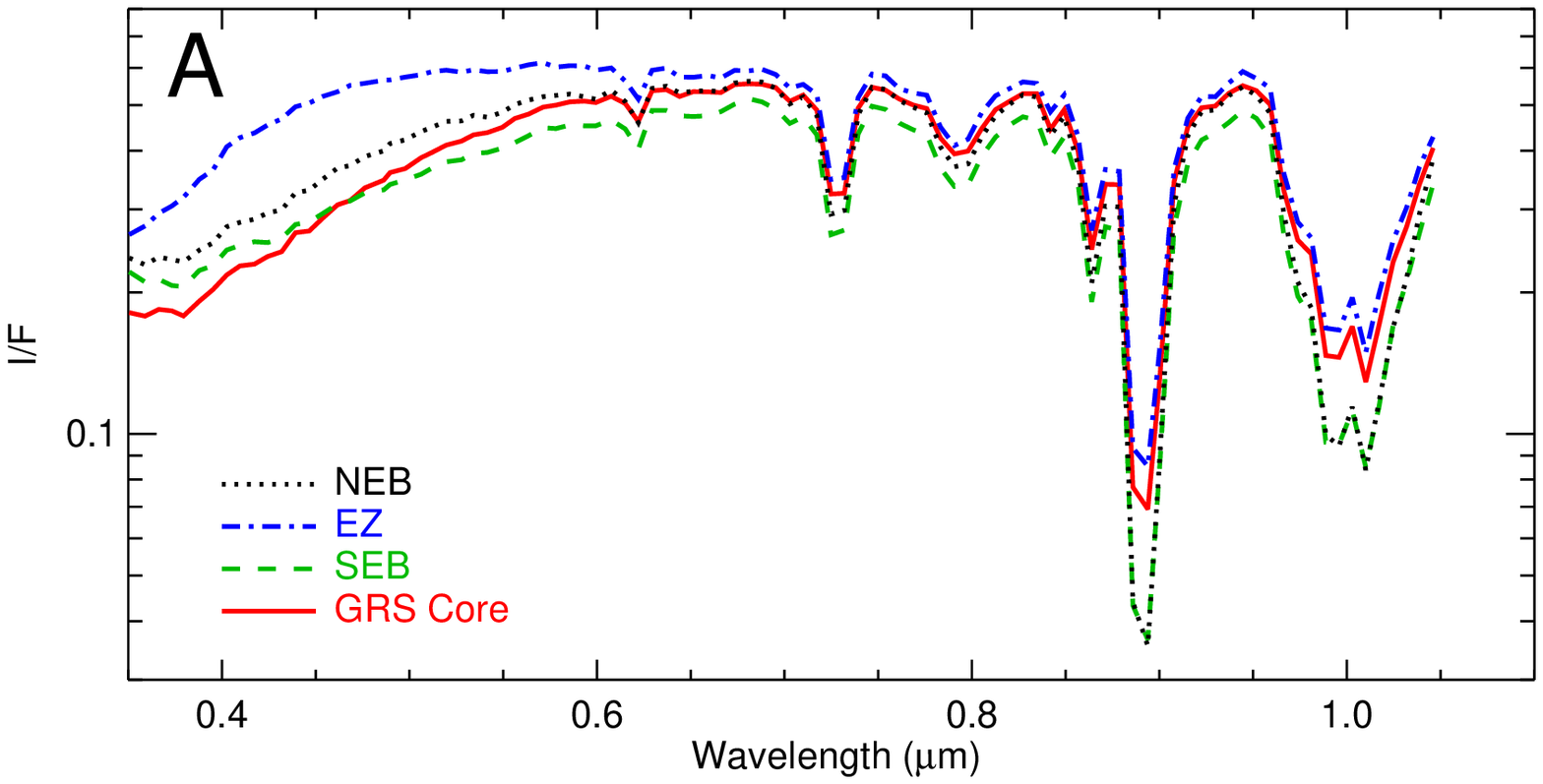}
\includegraphics[width=3.35in]{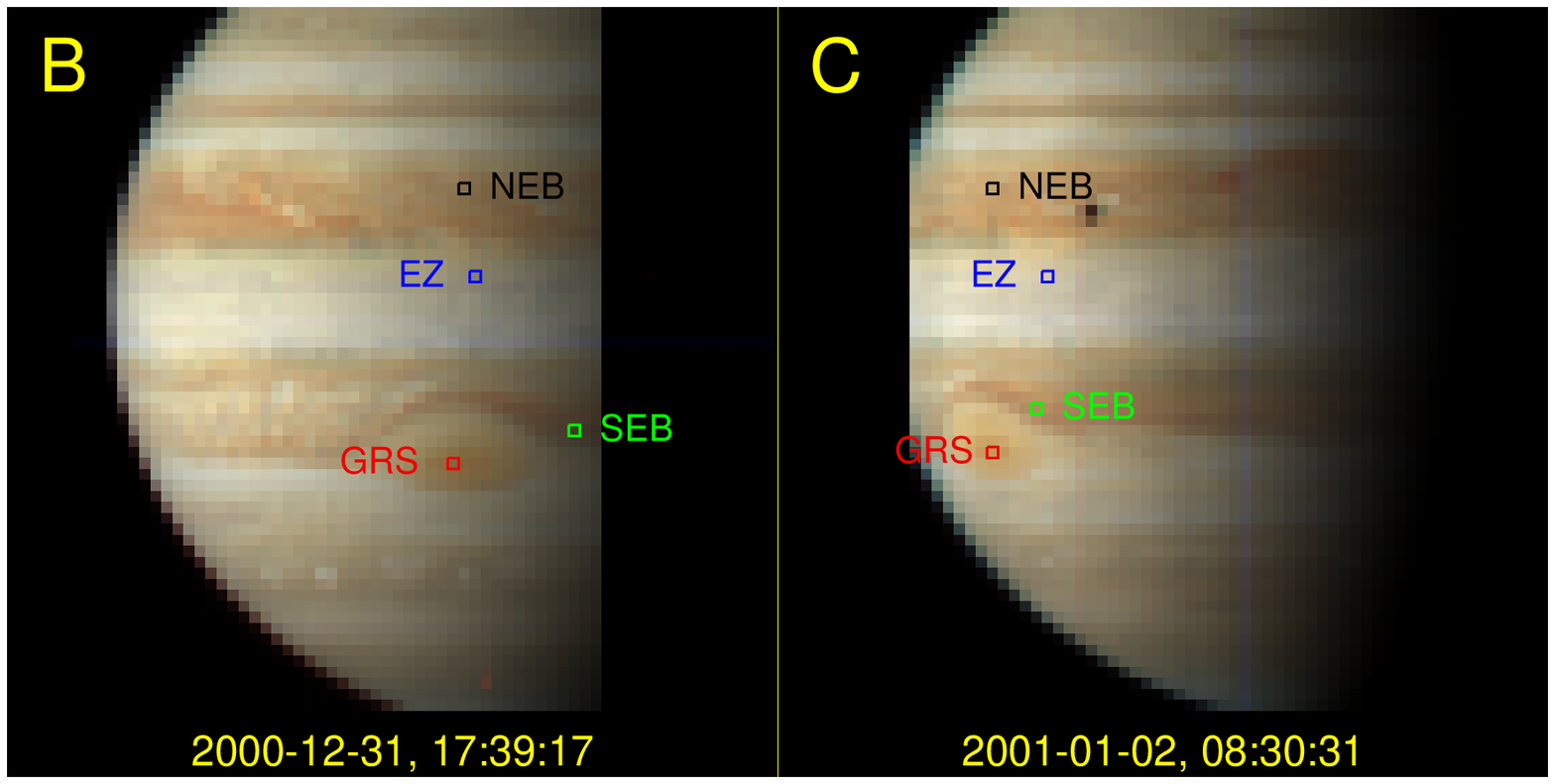}
\caption{A: Spectral samples from VIMS image cube V1357116132\_1 (from
  which the color composite in C was made). B: color composite image
  from VIMS cube V1356976257\_3, taken when features in C were
  positioned closer to the terminator. The wavelengths used for blue, green,
  and red channels in the composite images are 450, 550, and 750 nm,
  respectively.
  The locations of spectral samples are
  indicated by colored squares and labels.  Observation C was made
  38.85 hours after observation B. Samples in the second cube were
  acquired at positions as close as possible to the predicted
  locations of features shown in B, accounting for drift due to
  Jupiter's zonal wind profile.  The dark pixel in image C (below the B in NEB)
is due to the transit of Europa.}
\label{Fig:medphase}
\end{figure}

In December 2000, the Cassini spacecraft passed near Jupiter for a
gravitational assist on its journey to Saturn. During the flyby
it observed Jupiter using a suite of instruments that
included VIMS, which
provided spectral observations of Jupiter's atmosphere under
conditions summarized in Table\ \ref{Tbl:obslist}.  The spatial
resolution of the observations is limited by the rather large distance
of the flyby.  The low phase angle image acquired on 4 December is
well suited for comparison with groundbased observations that provide
a verification of the VIMS calibration at CCD wavelengths, but has
relatively low spatial resolution.  Images extracted from this data
set and spectral samples at key locations can be seen in Fig.\ \ref{Fig:lowphase}.
The two observations at intermediate phase angles, from 31 December and 2 January,
have much better spatial resolutions and
provide two different observing geometries of the GRS and
other cloud features, yielding additional constraints on
radiative transfer models.  Example images and spectral
samples from these later data are provide in Fig.\ \ref{Fig:medphase}

\begin{table*}\centering
\caption{Summary of VIMS observations used in our analysis.}
\vspace{0.15in}
\begin{tabular}{r c c c}
VIMS-VIS Cube:  & V1354610545\_3 & V1356976257\_3 & V1357116132\_1\\[0.1ex]
\hline
Date & 2000-12-04 & 2000-12-31 & 2001-01-02\\[0.1ex]
Time (UT) & 08:31:02 & 17:39:17 & 08:30:31\\[0.1ex]
Obs. Name:  {\small VIMS\_C23JU\_+} & {\small 6ATM2X2163\_ISS} & {\small ATMOS02A000\_ISS} & {\small FEATURE005\_CIRS}\\[0.1ex]
Sampling Mode & HI-RES & HI-RES & HI-RES\\[0.1ex]
Sun-Jupiter Distance (AU) & 5.0379 & 5.0460 & 5.0464\\[0.1ex]
Phase Angle & 9.76\deg & 67.83\deg & 76.77\deg\\[0.1ex]
S/C-Jupiter Range (km) & 26,439,177  &  9,873,972  & 10,185,177 \\[0.1ex]
Sub-observer Lat (centric) & 3.626\deg & -0.283\deg & -0.869\deg\\[0.1ex]
Sub-observer Lon (east) & 318.506\deg & 321.962\deg & 1.719\deg\\[0.1ex]
Sub-solar Lat (centric) & 2.948\deg & 2.901\deg & 2.898\deg \\[0.1ex]
Sub-solar Lon (east) & 328.254\deg & 254.185\deg & 285.016\deg  \\[0.1ex]
GRS emission cosine & 0.8878 & 0.8766 & 0.48420\\[0.1ex]
GRS incidence cosine & 0.8648 & 0.5707 & 0.8518\\[0.1ex]
GRS azimuth & 160.48\deg & 71.61\deg & 65.67\deg \\[0.1ex]
\hline
\end{tabular}\label{Tbl:obslist}
\end{table*}

\subsection{VIMS Calibration and Navigation}

\begin{figure}\centering
\includegraphics[width=3.45in]{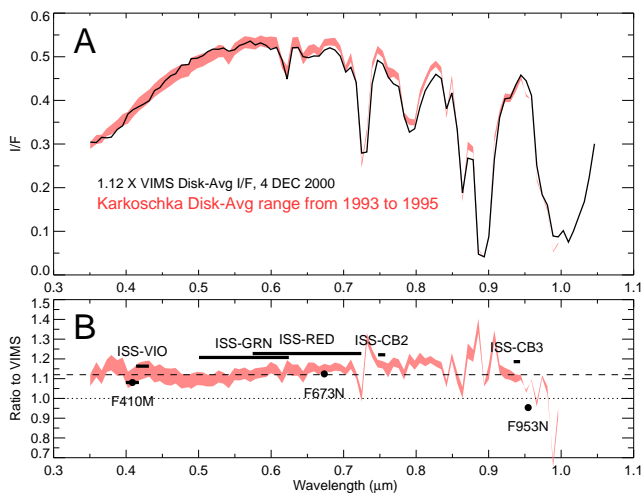}
\caption{{\bf A:} Disk-averaged I/F of Jupiter based on VIMS observations in
  December 2000 (solid curve, which is scaled by a factor of 1.12)
  compared to groundbased measurements in July 1993
  \citep{Kark1994Icar} and July 1995 \citep{Kark1998Icar} which are
  shown by a red band covering the range of the two groundbased
  measurements. {\bf B:} Ratios to the VIMS spectrum of Karkoschka's
  spectra, and to 14 December 1999 HST band-pass filter observations
  using F410M, F673N, and F953N filters, taken at a phase angle of
  11.46\degx, and ISS band-pass filter images obtained within one hour
  of the VIMS observations.  Karkoschka data were obtained at the
  European Southern Observatory, and Jupiter phase angles were 9.8\deg
  and 6.8\degx, respectively. Reference lines at ratios of 1.0 (dotted) and 1.12
(dashed) are also shown.  
}\label{Fig:diskint}
\end{figure}

The VIMS data set we used was reduced and calibrated using the USGS
ISIS3 \citep{Anderson2004} vimscal program, which was derived from the
software provided by the VIMS team (and is available on PDS archive
volumes), and uses the same calibration files and solar spectrum.  A
sanity check on the calibration was obtained by computing a disk
average spectrum from the low-phase angle VIMS cube and comparing it
with disk-averaged observations of \cite{Kark1998Icar} taken in 1995
and 1993.  The results, shown in Fig.\ \ref{Fig:diskint}, are
plausibly consistent within 10\%, given the the time difference
between the two observations sets.  A more contemporaneous calibration
check is provided by Cassini ISS band-pass filtered images of the type
shown in Fig.\ \ref{Fig:lowphase}B, which provide comparisons at
discrete wavelength bands, as shown in Fig.\ \ref{Fig:diskint}B.  The
ISS calibration leads to I/F values that are 20\% greater than
produced with the VIMS calibration. Given that the ISS calibration is
being revised to put it into better agreement with the Karkoschka groundbased
observations, we also considered Hubble Space Telescope WFPC2 observations obtained
on 14 October 1999 as another
sanity check.  Unfortunately, most of those WFPC2 observations were overexposed,
partially saturated, and not usable for disk-integrated comparisons.  Three
images that were not saturated (using filters F410M, F673N, and F953N) produced
disk-integrated I/F values that are compared with VIMS in Fig.\ \ref{Fig:diskint}B.
Given that the F410M and F673N images have the most reliable calibration, we interpret
these results to mean that the HST WFPC2 calibration leads to I/F values about
10\% higher than the VIMS results, which is close to the 12\% higher values we
estimated for the Karkoschka groundbased calibration.  As the HST results
are only about 14 months earlier than the VIMS observations in question,
while the Karkoschka measurements were 5 and 7 years earlier, that they
lead to essentially the same disk-averaged results suggests that Jupiter's
disk-averaged reflectivity is likely stable within about the 4\% uncertainty
claimed for the Karkoschka measurements.  The net impact of these comparisons
is that we should seriously consider raising the VIMS I/F by about 10\%.

The navigation of the ISIS3-processed cubes had to be adjusted to correct pointing errors. That was done
using limb fitting and cross checked with contemporaneous ISS wide-angle imaging observations.

\section{Radiative Transfer Modeling}

\subsection{Atmospheric structure and composition}

We used the tabulated results of \cite{Seiff1998} for Jupiter's
temperature structure down to the 22-bar level, and assumed a dry
adiabatic extrapolation below 22 bars.  We assumed an atmospheric
composition of He/H$_2$=0.157$\pm$0.003 \citep{VonZahn1998} and
CH$_4$/H$_2$= 2.1$\times 10^{-3}$$\pm$0.4$\times 10^{-3}$ \citep{Niemann1998}, which are
expressed as number density ratios.  Because \nht is a condensable gas
and is vertically variable below the condensation level, as well as
horizontally variable, we selected a parameterized profile that fit
the observed spectrum.  The parameterization is described in Sec.\ \ref{Sec:nh3par}.
We assumed that the hydrogen
para fraction matched the local equilibrium value.

\subsection{\nht parameterization}\label{Sec:nh3par}
\begin{figure*}[!htb]\centering
\includegraphics[width=3.in]{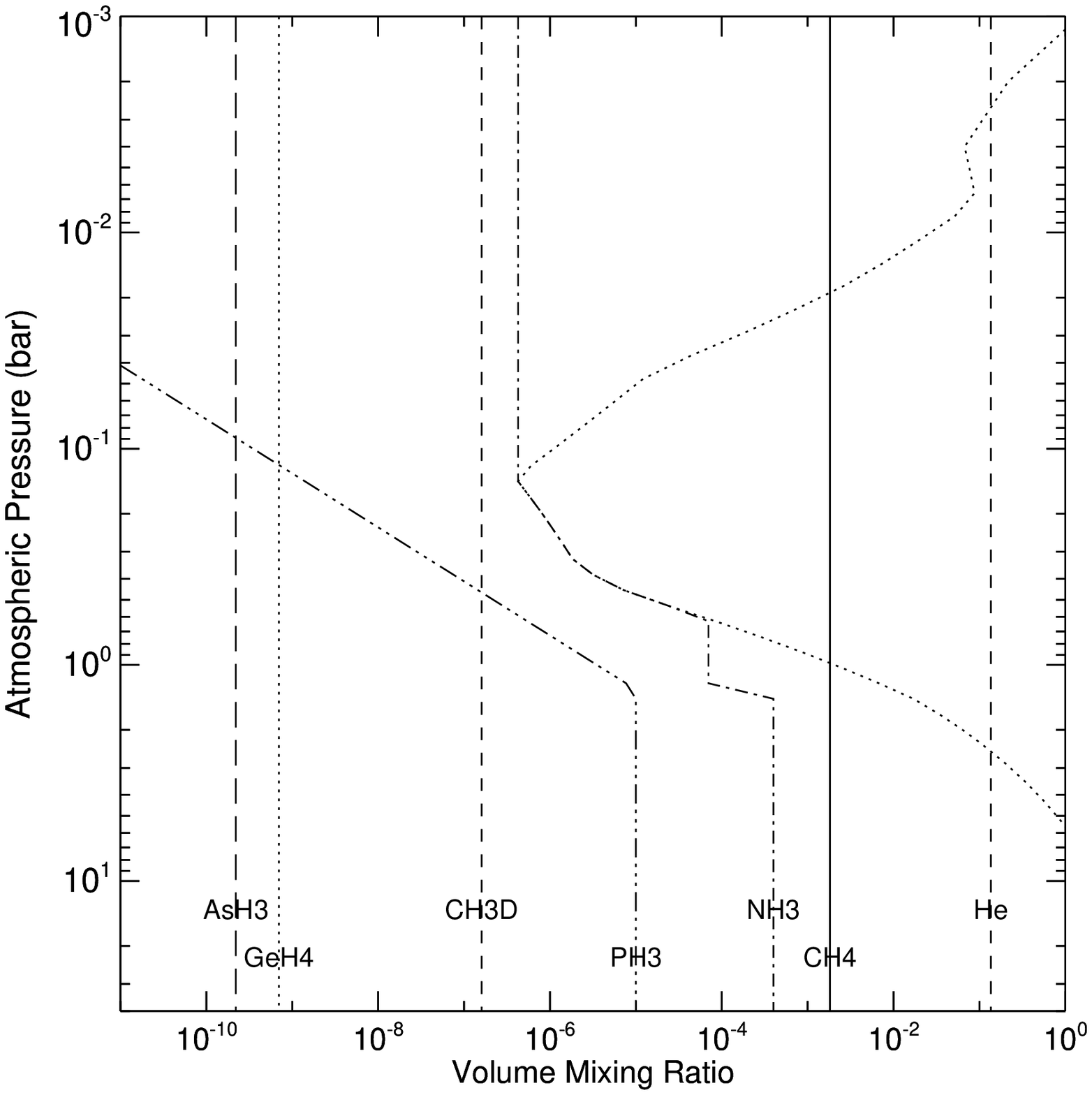}
\includegraphics[width=3.2in]{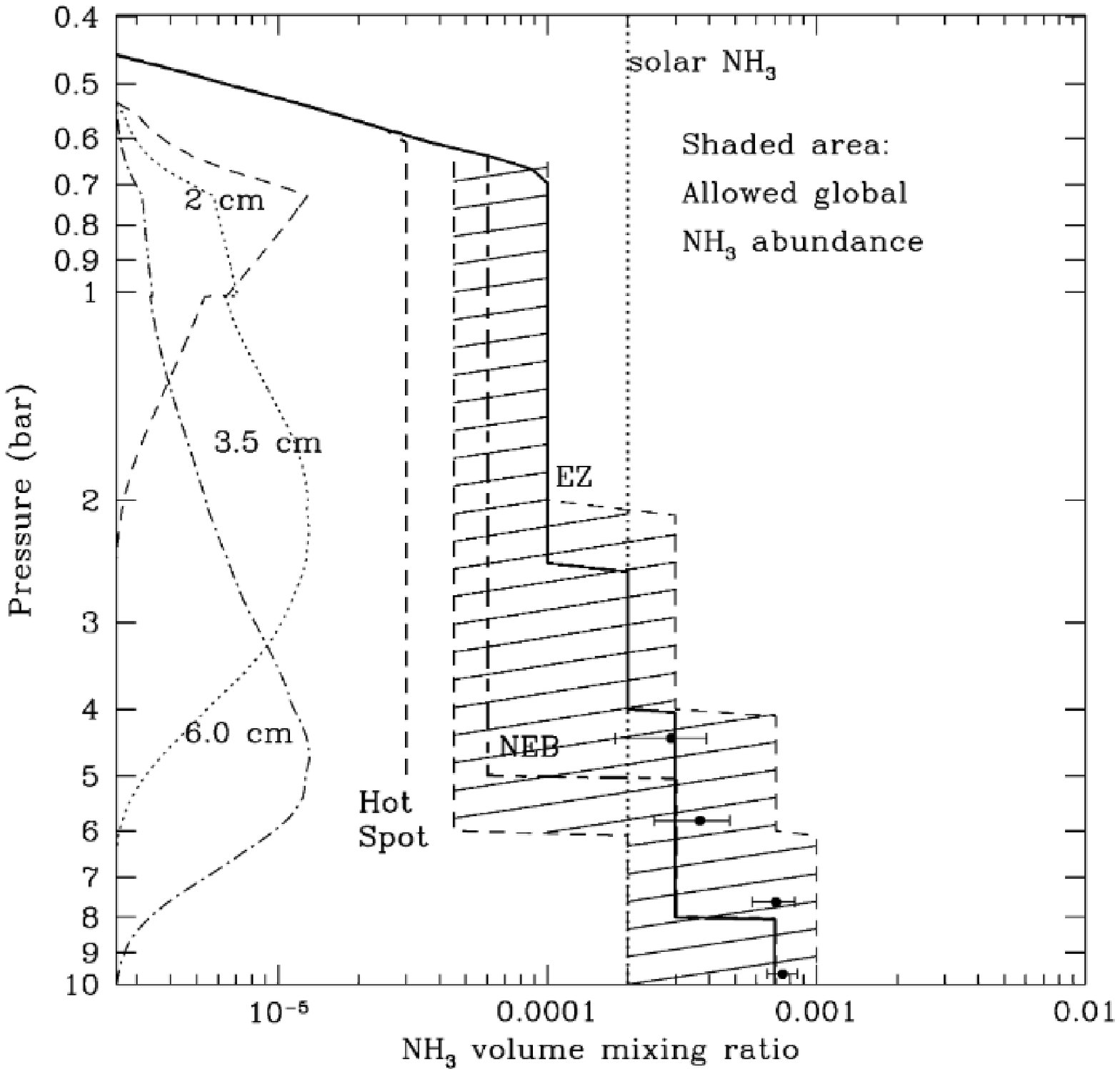}
\caption{Left: Vertical gas profiles assumed in radiation transfer
  modeling. Only \nhtx, \chfx, and H$_2$ have significant effects in
  the CCD spectral range, as indicated in
  Fig.\ \ref{Fig:pendepth}. The \nht profile is defined by adjustable
  parameters constrained to best fit \nht spectral features. Right:
  \nht volume mixing ratios consistent with radio observations, from
  \cite{Showman2005}. The plotted points are from
  \cite{Folkner1998}.\label{Fig:composition}}
\end{figure*}

We characterized the \nht profile using three parameters listed in
Table\ \ref{Tbl:paramlist}: a pressure break point $p_1$, a deep
mixing ratio $nh_3v_0$ for $p>p_1$, and a depleted mixing ratio
$nh_3v_1$ for $p_c < p < p_1$, where $p_c$ is the condensation
pressure.  A sample profile is shown in the left panel of
Fig.\ \ref{Fig:composition}, which also displays profiles consistent
with radio observations in the right panel.  We assumed that \nht is
saturated above the condensation level, an arbitrary choice to which
the visible spectra are not very sensitive. There is insufficient
sensitivity to \nht in the CCD spectral range to allow determination
of all three profile parameters independently.  It is well established
that the \nht mixing ratio increases with depth, based on radio
observations \citep{DePater1986Icar,DePater2016Sci}, as well as
Galileo probe observations \citep{Sro1998,Folkner1998}. What is less
well defined is the nature of the transition between deep and upper
tropospheric values, although it seems clear that belts are more
depleted than zones and likely depleted to somewhat greater depths
\citep{DePater1986Icar}, as indicated in the right panel of
Fig.\ \ref{Fig:composition}.  In view of this characteristic we
decided to use a fixed value of 4$\times 10^{-4}$ for the deep mixing
ratio, which is close to the Folkner et al. value at 6 bars
(Fig.\ \ref{Fig:composition}) and well within the range allowed by
radio observations, and make the upper mixing ratio and transition
pressure adjustable parameters constrained to minimize \chisqx.
However, given the generally weak effect of these parameters on the
observed CCD spectrum, these parameters cannot be tightly constrained.

\subsection{Gas absorption models}\label{Sec:gasabs}  
\vspace{-0.05in}

\begin{figure*}\centering
\includegraphics[width=5.in]{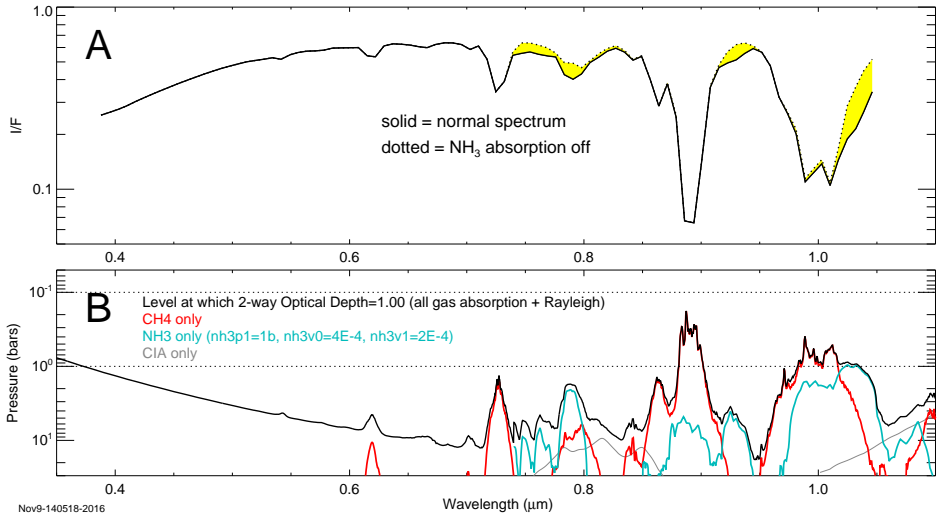}
\caption{A: Model spectra with \nht absorption included (solid) and
  without (dotted), with the difference colored yellow.  B: Pressure
  at which 2-way optical depth reaches unity vs wavelength for
  individual gases and for all gases combined, assuming a methane
  mixing ratio of CH$_4$/H$_2$= 2.1$\times 10^{-3}$ the ammonia
  profile parameters are given in the legend.}
\label{Fig:pendepth}
\end{figure*}

The importance of the various gas absorbers is indicated by the
penetration depth profile in Fig.\ \ref{Fig:pendepth}.  The two-way unit
optical depth level is shown individually for absorptions by methane
(red), ammonia (cyan), and collision-induced absorption (gray).  These
are compared to the unit extinction optical depth level (2-way) for
all gas absorptions combined together with Rayleigh scattering
extinction.  Methane and ammonia are the dominant gases that
shape the features in Jupiter's visible (CCD) spectrum. The lesser
role of ammonia absorption is illustrated by the comparison of model
spectra with and without ammonia absorption that is shown in the top panel
of Fig.\ \ref{Fig:pendepth}.

Gas absorptions are all modeled using 10-term correlated-k fits. For
methane these are based on band models published by \cite{Kark2010ch4}
(we used P. Irwin's fits, which are available at
http://users.ox.ac.uk/$\sim$atmp0035/ktables/ in compressed files
ch4\_karkoschka\_IR.par.gz and ch4\_karkoschka\_vis.par.gz).  Ammonia
models are based primarily on \cite{Bowles2008} (in this case we used
the fit described by \cite{Sro2010iso}). Collision-induced absorption
(CIA) for H$_2$ and H$_2$-He was calculated using programs downloaded
from the Atmospheres Node of the Planetary Data System, which are
documented by \cite{Borysow1991h2h2f, Borysow1993errat} for the
H$_2$-H$_2$ fundamental band, \cite{Zheng1995h2h2o1} for the first
H$_2$-H$_2$ overtone band, \cite{Brodbeck1999} for the second
H$_2$-H$_2$ overtone band, and by \cite{Borysow1992h2he} for H$_2$-He
bands.
%
%
  Where \chf
and \nht gas absorptions overlap we compute opacities for 100
combinations of 10 \chf terms by 10 \nht terms and then sort and refit
to a 10-term weighted sum.

\subsection{Radiation Transfer Code}

We used radiation transfer code based on that described by
\cite{Sro2010iso} and \cite{Sro2010vims}, with some enhancements and
exceptions. The IR code was extended to CCD wavelengths so that
correlated-k models could be used at short wavelengths.  Increased
parallelism was added so that wavelengths as well as correlated-k
terms could be run in parallel.  
Our two-component coated-sphere particle scattering code is based on
algorithms originated by \cite{Toon1981}.
 We approximated the line-spread function of the VIMS instrument as a
 Gaussian of FWHM = 0.007 $\mu$m, then collected all the opacity
 values within $\pm$FWHM of the sample wavelength, weighted those
 according to the relative amplitude of the line-spread function, then
 sorted and refit to ten terms again.  

We ignored both Raman scattering and polarization effects, based on
comparing trial calculations including these effects with those that
did not.  For the \cbm model, Raman scattering had no noticeable
effects beyond 370 nm, and including polarization depressed the I/F by
a mean of only 0.8\% and an RMS deviation of only 0.3\%, both well
below our estimated uncertainties.  We used a model atmosphere with 57
layers between 0.5 mbar and 40 bars, with additional layers introduced
to account for cloud pressure boundaries and the ammonia pressure
break point appearing between our initial layer boundaries.  We used 16
zenith angle quadrature points per hemisphere and 16 to handle
azimuthal variations. Calculations with 12 quadrature points in each
dimension had a maximum difference of only 0.03\% of I/F, implying
that 16 quadrature points per hemisphere were more than adequate.

\section{Parameterization of cloud structure and optical properties}

\subsection{Chromophore optical properties}

\begin{figure}\centering
\includegraphics[width=3.5in]{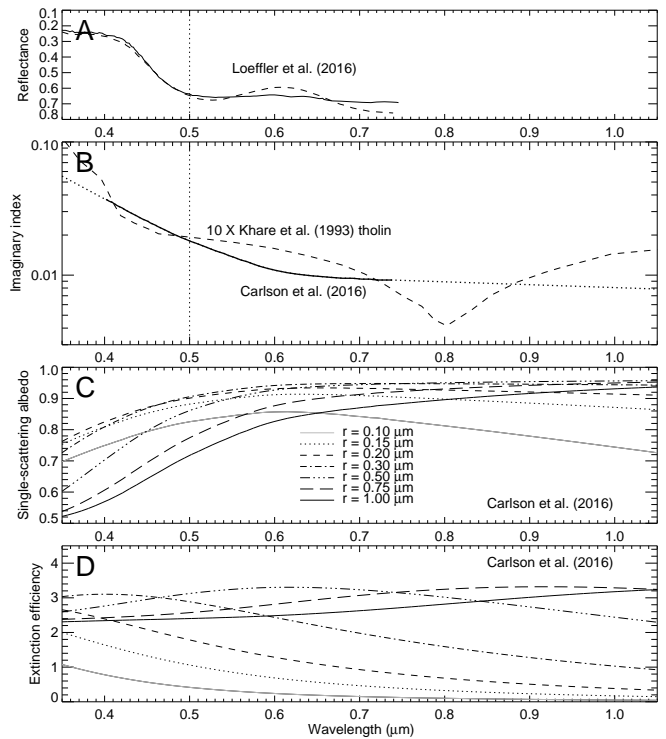}
\caption{Chromophore imaginary index (B), single-scattering albedo (C),
and extinction efficiency (D), 
all for particle sizes from 0.1 \mum to 1.0 \mumx, assuming
a gamma size distribution with $b=0.1$. In (B) the \cite{Carlson2016}
 imaginary index measurement is displayed by the thick solid line and the
assumed log-linear extrapolations are shown by the dotted line. The
dashed line displays the tholin index of \cite{Khare1993} multiplied by
a factor of ten. For reference, panel A displays the \cite{Loeffler2016}
measurement of reflectivity of \nhfsh irradiated 
at a temperature of 120 K and warmed to a temperature of 190 K (dashed)
and 200 K (solid). The y-axis scale in A is inverted so that increasing
absorption is in the same sense (upward) as the refractive index plots in panel B.}
\label{Fig:scatprop}
\end{figure}

The imaginary index and scattering properties of the
\cite{Carlson2016} chromophore are displayed in
Fig.\ \ref{Fig:scatprop}.  This material is nearly an order of
magnitude more absorbing than the tholin measured by \cite{Khare1993},
also shown for comparison.  The imaginary index of the new material also has a nearly
constant logarithmic slope over the 0.4-0.6 \mum wavelength range, which turns out
to be a desirable feature in matching the visible spectra of Jovian
clouds. Note the large change in slope of the tholin index in this
range.  Although \cite{Loeffler2016} did not measure the refractive
index of the material they produced (created by irradiation of
\nhfshx), the reflectivity they measured (reproduced in panel A)
contains substantial slope changes at 0.43 \mum and 0.5 \mum that are
not typical of the Jovian cloud spectra shown in
Figs.\ \ref{Fig:lowphase} and \ref{Fig:medphase}. The scattering
properties of pure chromophore particles, shown in panels B-D, assume
a real index of $n$=1.4, following \cite{Carlson2016}.  These panels
show imaginary index, single-scattering albedo, and extinction
efficiency, all for particles made entirely of chromophore material
and for particle radii from 0.1 to 1.0 \mumx. In panel B the measured
index is indicated using a thick solid line and the extrapolated
values indicated by a thin dotted line.  Note the dramatic spectral
variations possible in scattering parameters obtained by changes in
particle radius.  This shows that small changes in particle size can
have a dramatic impact on the color of light transmitted by a layer of
such particles.

\subsection{Cloud model structure}

The cloud model we used to reproduce the observed VIMS visual channel
spectra is a relatively simple one, containing three layers of
Mie-scattering spherical particles.  Our parameterization is
summarized in Table\ \ref{Tbl:paramlist}. The particles were assumed
to have gamma size distributions \citep{Hansen1974} with variance parameter $b=0.1$.
The top layer is a stratospheric haze defined by
an effective pressure $p_1$, a 1-\mum optical depth $\tau_1$, an effective particle size
$r_1$, and a refractive index $n_1(\lambda)$.  This layer is treated as a sheet
cloud arbitrarily placed at 40 mbar.  Its effect is found to be too small to
reasonably constrain its vertical distribution.
The next layer is the main cloud, which is
parameterized by top pressure $p_{2\mathrm{T}}$ and bottom pressure $p_2$, a
particle radius $r_2$, refractive index $n_2(\lambda)$,
and optical depth $\tau_2$.  For what \cite{Baines2016Icarus} called the \cbm model, which
is the model we use in this analysis, there is a third layer
tacked onto the top of the main layer. This third layer contains the
red chromophore and is characterized by its optical depth $\tau_3$ and
particle radius $r_3$, as well as its refractive
index $n_3(\lambda)$, which is taken from \cite{Carlson2016}.

Because our model treats the main cloud layer as conservative, it is not an effective
barrier to 5-\mum emission from deeper layers of Jupiter's atmosphere.
While that layer is not important in shaping the CCD portion of Jupiter's
reflection spectrum, it leaves the emission blocking to some deeper cloud
layer that is not included here.  Alternatively, one could add a small amount of absorption to the
main cloud layer so that it could do much, or even all, of the thermal
blocking. However, that absorption is not needed to fit the visible
spectrum, and finding a cloud structure that matches both thermal
and reflected sunlight is left for future work.

\begin{table*}\centering
\caption{Summary of cloud and \nht model parameters used in spectral calculations.}
\vspace{0.15in}
\begin{tabular}{r l l }
Name, unit & Description & Value\\
\hline
$p1$, bar & stratospheric haze pressure & 0.04-.07\\
$r1$, \mum & stratospheric haze particle radius & fixed or adjustable \\
$\tau_1$ & stratospheric haze optical depth at 1 \mum & adjustable\\
$n_1(\lambda)$ & stratospheric haze refractive index & $n_1=1.4+0i$, or Carlson et al. (2016)\\
$p_{2\mathrm{T}}$, bar & top of main cloud & adjustable\\
$p_2$, bar & bottom of main cloud & adjustable\\
$r_2$, \mum & effective radius of main cloud particles & adjustable\\
$\tau_2$ & optical depth of upper cloud at 1 \mum & adjustable\\
$n_2(\lambda)$ & refractive index of main cloud & $n_2=1.4+0i$ or $n_2=1.85+0i$\\
$p_{3T}$, bar & pressure at top of chromophore layer & normally = $p_{2T} \times 0.9$\\ 
$p_3$, bar & pressure at base of chromophore layer & normally = $p_{2T}$\\
$r_3$, \mum & effective radius of chromophore layer & adjustable\\
$\tau_3$ & optical depth of chromophore layer at 1 \mum & adjustable\\
$n_3(\lambda)$ & chromophore layer refractive index & Carlson et al. (2016)\\
$H_c/H_g$ & cloud particle to gas scale height ratio & normally = 1.0 \\
$nh_3v_0$ & \nht vmr for $p > nh_3p_1$ (deep mixing ratio) & set to $4\times10^{-4}$ \\
$nh_3v_1$ & \nht vmr for $nh_3p_1 > p >$ condensation $p$ & adjustable\\
$nh_3p_1$, bar & \nht break-point pressure & adjustable\\ 
\hline\\[-0.1in]
\end{tabular}
\parbox{5.2in}{NOTE: aerosol particles are assumed
  to have a gamma size distribution with variance parameter $b=0.1$,
  with distribution function $n(r) = \mbox{constant}\times
  r^{(1-3b)/b} e^{-r/ab}$, where with $a = r_{eff}$ and $b=$
  dimensionless variance, following \cite{Hansen1974}.  Our model does
  not include a deep absorbing cloud that is needed to block thermal
  emission in the 5-\mum region of the spectrum.}
\label{Tbl:paramlist}
\end{table*}

\subsection{Refractive index of main cloud layer}

The main cloud layer is a strong absorber near 3 \mumx, and is likely
composed of some mixture of \nht and \nhfsh
\citep{Sro2010iso,Sro2010vims}.  If dominated by \nhtx, the real
refractive index $n_2$ would be near 1.4, while if predominantly
\nhfshx, the main layer's index would be closer to 1.8.
\cite{Sato2013} found a best-fit real refractive index of 1.85 for
Jovian clouds in the equatorial zone, favoring the dominance of
\nhfsh in that location.  In our analysis we tried models with both $n_2=1.4$ and
$n_2=1.85$, finding that the \cite{Carlson2016} chromophore provides a
good fit to Jupiter's varied colors in either case.

\subsection{Vertical location of chromophore particles}

The thin chromophore layer of the \cbm model is bounded on the bottom
by the pressure at the top of the main cloud layer $p_{2\mathrm{T}}$ and on the top by
0.9$\times p_{2\mathrm{T}}$. The arbitrary thickness of 0.1$\times p_{2\mathrm{T}}$ ($\sim$20 mbar) is not
well constrained by the observations.  For the low phase-angle
observations, it appears that the chromophore could be in the
stratospheric haze, or on top of the main cloud, or as a diffuse haze
extending above the main cloud.  
This is consistent with the results of \cite{Baines2016Icarus}, who
considered two additional models: (1) models in which the chromophore
material was only in a stratospheric haze, and (2) models in which the
chromophore was a coating on the main cloud particles. Because neither
of those models produced fits as accurate as the \cbm model, we mainly
used the \cbm model in our analysis.  However, models with chromophores
in the stratosphere still provided relatively good fits to the spectrum
and it is not possible to rule out a vertically distributed chromophore.


\subsection{Sensitivity to model parameters.}\label{Sec:sens}  

\begin{figure}\centering
\includegraphics[width=3.5in]{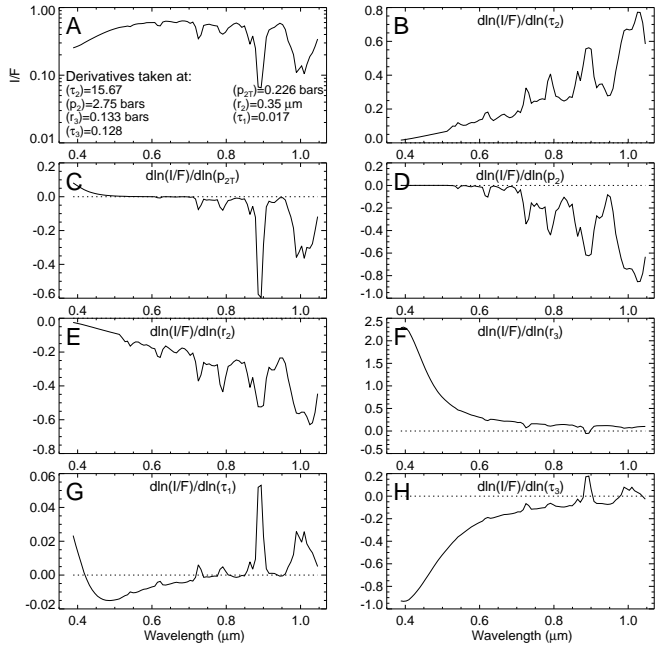}
\caption{I/F spectrum (A) and derivatives of fractional changes in I/F with respect
to fractional changes in parameters $\tau_2$ (B), $p_{2\mathrm{T}}$ (C), $p_2$ (D),
$r_2$ (E), $r_3$ (F), $\tau_1$ (G), and $\tau_3$ (H).}\label{Fig:deriv}
\end{figure}

\begin{figure}\centering
\includegraphics[width=3.5in]{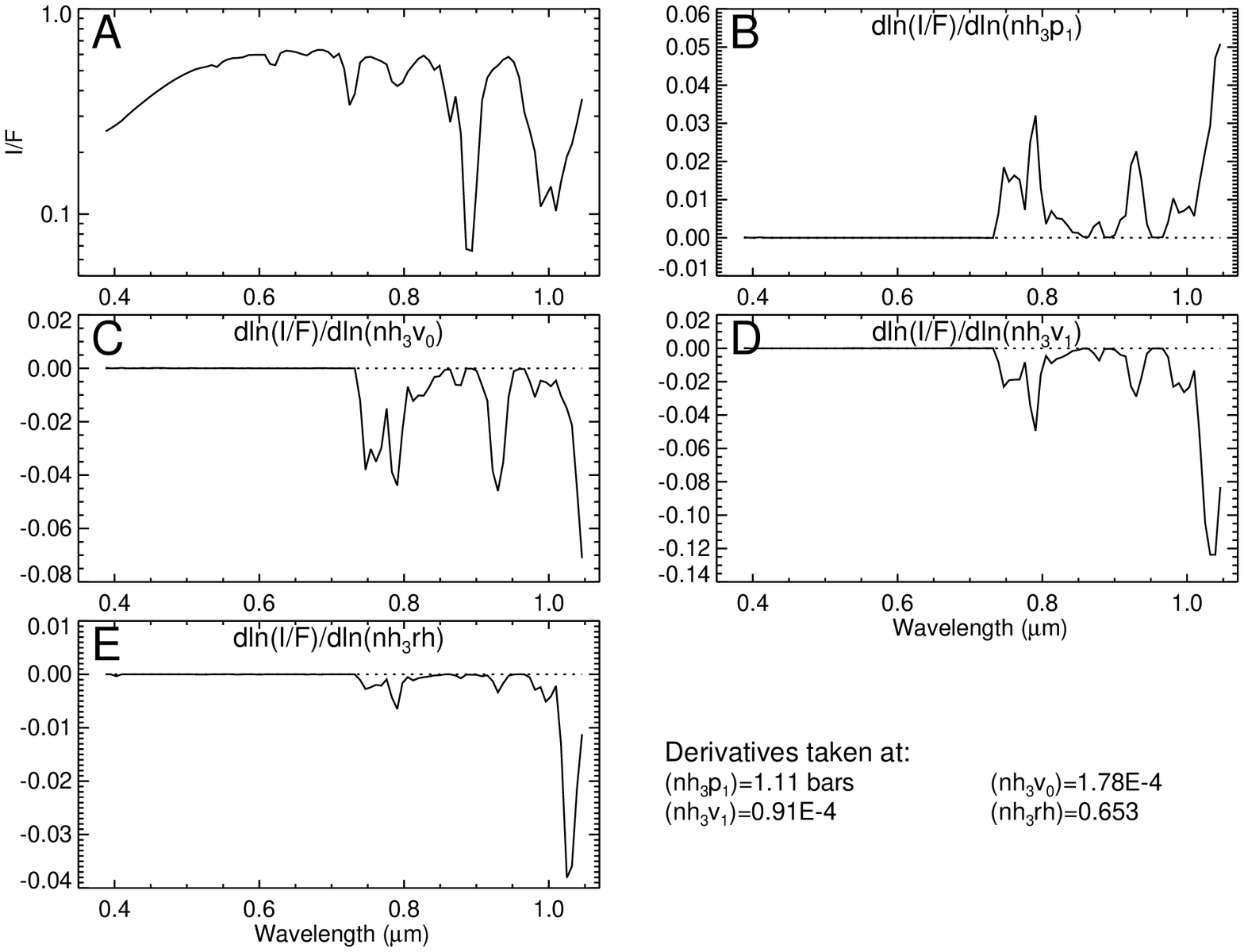}
\caption{I/F spectrum (A) and fractional changes in I/F with respect
  to fractional changes in parameters $nh_3p_1$ (B), $nh_3v_0$ (C), $nh_3v_1$ (D),
  and $nh_3rh$ (E), for conditions shown in the legend (from a model
  given in Table\ \ref{Tbl:lowphase} that fits the low phase angle GRS core
  spectrum).  These parameters generally have a much smaller effect
on the spectrum than those shown in Fig.\ \ref{Fig:deriv}, where a fractional
change in those parameters can produce roughly the same fractional change in
parts of the I/F spectrum. Here the fractional changes in I/F are typically
only 5-10\% of the fractional parameter changes.}\label{Fig:gasder}
\end{figure}

For the GRS \cbm model, we computed derivatives of the spectrum with
respect to the fitted parameters.  The results are shown in
Figs.\ \ref{Fig:deriv} and \ref{Fig:gasder} as fractional derivatives,
i.e. the fractional rate of change of I/F with respect to the
fractional change in each fitted parameter. For example, panel B of
Fig.\ \ref{Fig:deriv} shows that if the optical depth of the main
cloud is increased by 10\%, the I/F at 1 \mum would be increased by
about 8\%, but hardly changed at all at 0.4 \mumx, where Rayleigh
scattering provides a significant contribution. The importance of
Rayleigh scattering can also be seen in panel C, from the fact that
moving the top of the main cloud layer to higher pressure increases
the I/F at the shortest wavelengths.  In panel H, we see that a 10\%
increase in the optical depth of the chromophore layer would produce
about a 9\% decrease in I/F at 0.4 \mumx, while panel F shows that a
10\% increase in the radius of the chromophore particles would
produced more than double that increase in I/F at the shortest
wavelengths and almost nothing at the longest wavelengths.  

\begin{table*}\centering
\caption{Sample correlation matrix for cloud and gas adjustable parameters for
the \cbm model fit to the GRS spectrum shown in Fig.\ \ref{Fig:fitspecs}.}\label{Tbl:corr}
\vspace{0.1in}
\begin{tabular}{r | c c c c c c c c c} 
      &  $p_2$  &  $\tau_1$ &  $\tau_2$ & $r_2$  &  $p_{2\mathrm{T}}$ & $r_3$  &  $\tau_3$ &   $nh_3p_1$ & $ nh_3v_1$\\[.05in]
\hline\\[-0.05in]
$p_2$ & 1.0000 &-0.1265 & 0.4433 & 0.1450 &-0.3262 & 0.2556 & 0.2659 &-0.0617 &-0.2219\\[.05in]
$\tau_1$ &-0.1265 & 1.0000 & 0.3484 & 0.4322 & 0.8220 &-0.9265 &-0.9161 &-0.2114 &-0.1991\\[.05in]
$\tau_2$ & 0.4433 & 0.3485 & 1.0000 & 0.9394 & 0.2008 &-0.3994 &-0.4121 & 0.0040 & 0.2388\\[.05in]
$r_2$ & 0.1450 & 0.4323 & 0.9394 & 1.0000 & 0.2692 &-0.5463 &-0.5691 & 0.0929 & 0.3082\\[.05in]
$p_{2\mathrm{T}}$ &-0.3262 & 0.8220 & 0.2007 & 0.2692 & 1.0000 &-0.7311 &-0.7088 &-0.3587 &-0.1306\\[.05in]
$r_3$ & 0.2556 &-0.9265 &-0.3994 &-0.5463 &-0.7311 & 1.0000 & 0.9969 & 0.1405 & 0.0840\\[.05in]
$\tau_3$ & 0.2660 &-0.9161 &-0.4121 &-0.5691 &-0.7089 & 0.9969 & 1.0000 & 0.1196 & 0.0658\\[.05in]
$nh_3p_1$ &-0.0617 &-0.2114 & 0.0041 & 0.0930 &-0.3587 & 0.1405 & 0.1196 & 1.0000 & 0.4977\\[.05in]
$nh_3v_1$ &-0.2219 &-0.1991 & 0.2388 & 0.3082 &-0.1306 & 0.0840 & 0.0658 & 0.4977 & 1.0000\\[.05in]
\hline
\end{tabular}
\end{table*}

Note the strong similarity, but opposite sense, between the
derivatives of the bottom pressure of the main cloud
(Fig.\ \ref{Fig:deriv}D) and the optical depth of the cloud (panel B).
This arises because the most critical parameter controlling I/F is the
density of the cloud near its top (optical depth per bar). If the
optical depth is increased the I/F increases, more significantly at
the longer wavelengths, but if the bottom pressure increases while the
total optical depth is held fixed, then the optical depth per bar
decreases throughout the cloud and then so does the I/F. This means
that in fitting a spectrum, an increase in one requires an increase in
the other to compensate, leading to a positive correlation in the
fitting process, which has the value of 0.4433, as given in
Table \ref{Tbl:corr}.  The strongest correlation between fitted
parameters (0.997) is actually between $r_3$ and $\tau_3$. Note in panel C
the strong effect of the main cloud's top pressure on the I/F near
0.89 \mum and 1.0 \mumx, which implies that the observed I/F at these
wavelengths provide a strong constraint on the cloud top pressure.
The fractional derivatives with respect to the \nht profile parameters
(Fig.\ \ref{Fig:gasder}) are much weaker than most of the model
parameters, typically yielding only a few percent change in I/F for a
10\% parameter change.  Note that the two gas parameters we chose to
adjust have a roughly 50\% correlation.

\section{Results from fitting low phase angle VIMS observations.}

\subsection{Fit quality and parameter variations}

Our initial fits to the low phase angle observations of the GRS
showed that putting the chromophore in the stratospheric haze or at
the top of the main cloud produce comparable and excellent fits.  The
model in which the chromophore appears as a coating on the main cloud
particles provides an inferior fit.  With the chromophore as a coating
on all the particles in the main cloud, the UV-VIS reflectivity
gradient of the model can't reach the observed gradient without making
the longer wavelength continuum regions too dark.  This results in an
optimum core fraction of 0.96 of the total particle radius, or a
coating fraction 0.04. This fits within in the range of coated sphere
results given in the \cite{Carlson2016} paper, but does not yield a
good fit to the observed spectral gradient, mainly due to Rayleigh
scattering above the main cloud, which reduces the spectral gradient produced
at the main cloud level.

\begin{figure}\centering
\includegraphics[width=3.45in]{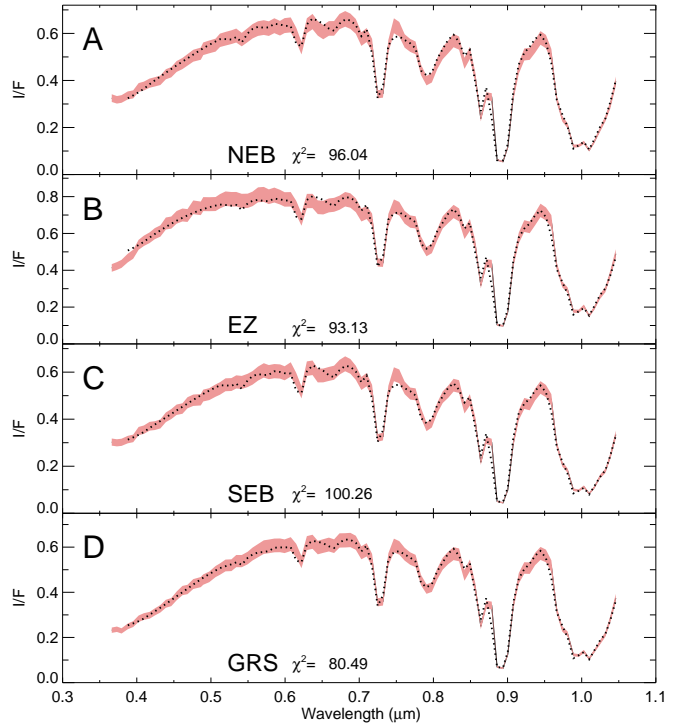}
\caption{\Cbm model spectra (dotted lines) fitted to low phase angle
  measured spectral samples from NEB (A), GRS (B), EZ (C), and SEB
  (D), shown as pale red bands bounding the assumed uncertainty range
  for combined model and measurement errors. The best-fit values of
  the adjusted parameters are given in Table\ \ref{Tbl:lowphase}.}
\label{Fig:fitspecs}
\end{figure}

The \cbm model fits to low phase angle spectra of GRS, SEB, EZ, and
NEB cloud structures are shown in Fig.\ \ref{Fig:fitspecs}, where
model spectra are shown as dotted lines in comparison with measured
spectra, displayed with a light red band covering the uncertainty
range assumed for measured and modeled spectra.
Numeric values of the fitted model parameters and their uncertainties are
given in Table\ \ref{Tbl:lowphase}. The fits are all roughly as good as
might be expected given estimated uncertainties.  For 91 wavelengths
and 9 fitted parameters, we should expect \chisq $\approx
N_\mathrm{F}$ = 91-9 = 82, within an uncertainty of $\sqrt{
  2N_\mathrm{F}}=13$.  Table\ \ref{Tbl:lowphase} shows reduced \chisq values
(\chisqx$/N_\mathrm{F}$) from 0.98 to 1.22, which satisfy these
expectations.

The main result of the fits for these very different cloud regions on Jupiter
is that the different degrees of reddish colors they exhibit can
all be accurately reproduced using the same chromophore described by \cite{Carlson2016},
when placed in a thin layer on top of the main cloud layer, changing only the optical depth
(from 0.13 for the GRS to 0.22 for the NEB) and the effective particle size (from 0.13 \mum for the GRS
to 0.33 \mum for the NEB).  {\it The Carlson et al. substance seems to be a nearly
universal chromophore for modeling Jupiter's colors.}  

\subsection{Chromophore total mass loading}

\begin{figure*}\centering
\includegraphics[width=5.2in]{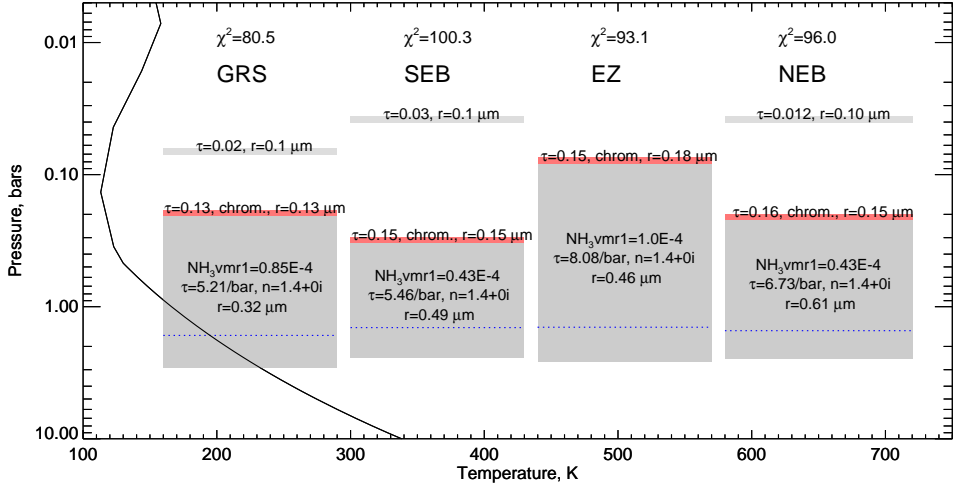}
\caption{\Cbm cloud model fits to low phase-angle VIMS spectral
  samples from NEB, GRS, EZ, and SEB regions on Jupiter.  The blue
  dotted line indicates the pressure below which the \nht vmr is
  assumed to be $4\times10^{-4}$.  Spectral fits are shown in
  Fig.\ \ref{Fig:fitspecs}. Adjusted parameter values and
  uncertainties are given in Table\ \ref{Tbl:lowphase}.}
\label{Fig:cartoon}
\end{figure*}

It is of some interest for models of production and evolution of the
chromophore material to estimate the total mass of chromophore present
per unit area in each of these cloud systems.  Assuming a density of
1 g/cm$^3$, and a single particle size instead of a distribution, the
mass per unit area can be estimated as the number of particles per
unit area times the volume of each particle. As given in
Table\ \ref{Tbl:lowphase}, the chromophore mass densities in units of
$\mu$g/cm$^2$ are 19 for the GRS, 18 for the SEB, 13 for the EZ,
and 20 for the NEB. This is a remarkable degree of uniformity in the
vertically integrated amount of coloring agent over such a wide range
of vertical cloud structures, as illustrated in
Fig.\ \ref{Fig:cartoon}. If the chromophore is produced by a chemical
process beginning with photolysis of ammonia, followed by chemical
reactions with acetylene, the production must be occurring at least
initially well above the cloud tops, where acetylene and ammonia
abundances produce the maximum reaction rate.  One would have thought
that a production at low pressures, then the process of coagulation
and sedimentation would result in a vertically distributed haze,
rather than what seems to be the case here, namely a haze of very
small scale height, plastered on top of the main cloud layer in each
region.  However, it is only for the GRS that we tested the
possibility of an extended haze. Perhaps for the NEB and SEB, we might
find a different result. The distributed haze was also not tested for
scaled up I/F values, nor for high-index main cloud particles.

\begin{table*}\centering
\caption{Best-fit parameter values for \cbm model fits to VIMS low phase-angle observations, using $nh_3p_1$ and  $nh_3v_1$ as adjustable \nht profile parameters.}
\begin{tabular}{ r c c c c}
Parameter, unit &     GRS (20.5\deg S) &     SEB (12.9\deg S) &       EZ (1.8\deg N) &     NEB (12.6\deg N)\\[0.05in]
\hline\\[-0.05in]
               $\tau_1$ & 0.020$^{+0.087}_{-0.017}$ & 0.027$^{+0.019}_{-0.012}$ & 0.000$^{+0.000}_{-0.000}$ & 0.012$^{+0.076}_{-0.011}$\\[0.05in]
          $p_{2\mathrm{T}}$, bar & 0.207$^{+0.043}_{-0.036}$ & 0.330$^{+0.045}_{-0.042}$ & 0.083$^{+0.031}_{-0.016}$ & 0.222$^{+0.049}_{-0.041}$\\[0.05in]
           $p_2$, bar & 2.923$^{+0.343}_{-0.311}$ & 2.418$^{+0.196}_{-0.179}$ & 2.624$^{+0.270}_{-0.244}$ & 2.465$^{+0.209}_{-0.191}$\\[0.05in]
               $\tau_2$ & 14.15$^{+ 4.54}_{- 3.48}$ &11.407$^{+2.141}_{-1.813}$ &20.520$^{+4.420}_{-3.683}$ &15.109$^{+1.786}_{-1.606}$\\[0.05in]
$\tau_2/(p_2-p_{2\mathrm{T}})$, optical depth/bar &    5.21 &   5.46 &   8.07 &   6.74\\[0.05in]
          $r_2$, \mum & 0.321$^{+0.101}_{-0.065}$ & 0.487$^{+0.132}_{-0.098}$ & 0.463$^{+0.100}_{-0.078}$ & 0.611$^{+0.129}_{-0.105}$\\[0.05in]
          $r_3$, \mum & 0.133$^{+0.016}_{-0.014}$ & 0.145$^{+0.013}_{-0.012}$ & 0.180$^{+0.019}_{-0.018}$ & 0.217$^{+0.013}_{-0.013}$\\[0.05in]
               $\tau_3$ & 0.128$^{+0.035}_{-0.028}$ & 0.146$^{+0.031}_{-0.026}$ & 0.154$^{+0.000}_{-0.000}$ & 0.330$^{+0.050}_{-0.046}$\\[0.05in]
$\tau_3/(\pi r_3^2 Q_\mathrm{ext})$, part./cm$^2$ &1.90$\times 10^9$ & 1.40$\times 10^9$ & 5.23$\times10^8$ & 4.67$\times10^8$\\[0.05in]
mass density, $\mu$g/cm$^2$ &  18.81 &  18.04 &  12.82 &  20.03\\[0.05in]
        $nh_3p_1$, bar & 1.650$^{+1.170}_{-0.590}$ & 1.440$^{+0.410}_{-0.280}$ & 1.430$^{+1.240}_{-0.508}$ & 1.520$^{+0.520}_{-0.340}$\\[0.05in]
$nh_3v_1$$\times 10^4$ & 0.849$^{+0.301}_{-0.235}$ & 0.432$^{+0.143}_{-0.110}$ & 1.020$^{+0.520}_{-0.393}$ & 0.426$^{+0.177}_{-0.128}$\\[0.05in]
$\chi^2$ &  80.49 & 100.26 &  93.13 &  96.04\\[0.05in]
$\chi^2/N$ &   0.98 &   1.22 &   1.14 &   1.17\\[0.05in]
\hline\\[-0.1in]
\end{tabular}\label{Tbl:lowphase}
\parbox{5in}{Note: Latitudes in column headings are planetocentric. The \nht  mixing ratio was set to $nh_3v_0=4\times 10^{-4}$
for $p>p1$.}
\end{table*}

\subsection{Effects of calibration and main cloud index variations}

Table\ \ref{Tbl:lowphasevar} summarizes the effects on fit-derived
parameters for the NEB region due to changing the refractive index of the main cloud
layer from 1.4 to 1.85, and the effects of multiplying the VIMS I/F
observations by a factor of 1.12 before doing the model fitting.  
As evident from the reduced \chisq values for these various cases,
the effects on fit quality are minimal and differ so little from the
nominal case that they do not warrant a comparison of the model spectra.
No matter which value of $n_2$ is chosen or which I/F calibration is chosen,
a model structure can be found that yields an excellent fit to the observed
spectra.  Thus our conclusions that the \cite{Carlson2016} chromophore can
be used to reproduce nearly any Jovian spectrum remains unaffected.
The main effect of increasing the observed I/F by a factor of 1.12 is to produce
modest increases in $p_2$, and $\tau_2$, slight decreases in $p_{2\mathrm{T}}$, $r_2$, $\tau_3$, and $nh_3p_1$,
with almost no change in the remaining parameters.  The refractive index
changes are more significant: mainly a nearly 50\% drop in $\tau_2$
and a substantial decease in $r_2$, with only modest decreases in $nh_3v_1$.
The fit quality differences are so slight as to provide no preference for
either calibration or refractive index options.

\begin{table*}\centering
\caption{Best-fit parameter values for \cbm model fits to VIMS low phase-angle observations, with different
choices for scale factor and $n_2(\lambda)$.}
\begin{tabular}{ r c c c c}
                &   $n_2=1.4+0i$ &  $n_2=1.4+0i$ &  $n_2=1.85+0i$ & $n_2=1.85+0i$\\[0.01in]
Parameter, unit &   scale=1.0 &  scale=1.12 & scale=1.0 & scale=1.12\\[0.05in]
\hline\\[-0.05in]
 $\tau_1$ & 0.012$^{+0.076}_{-0.011}$ & 0.000$^{+0.000}_{-0.000}$ & 0.000$^{+0.000}_{-0.000}$ & 0.000$^{+0.000}_{-0.000}$\\[0.05in]
   $p_{2\mathrm{T}}$, bar & 0.222$^{+0.049}_{-0.041}$ & 0.196$^{+0.029}_{-0.025}$ & 0.233$^{+0.030}_{-0.027}$ & 0.228$^{+0.030}_{-0.027}$\\[0.05in]
    $p_2$, bar & 2.465$^{+0.209}_{-0.191}$ & 2.825$^{+0.268}_{-0.246}$ & 2.624$^{+0.225}_{-0.207}$ & 2.993$^{+0.298}_{-0.276}$\\[0.05in]
 $\tau_2$ & 15.11$^{+ 1.79}_{- 1.61}$ &17.390$^{+3.803}_{-3.153}$ & 6.809$^{+0.436}_{-0.410}$ & 9.680$^{+0.923}_{-0.844}$\\[0.05in]
$\tau_2/(p_2-p_{2\mathrm{T}})$, optical depth/bar &    6.74 &   6.61 &   2.85 &   3.50\\[0.05in]
   $r_2$, \mum & 0.611$^{+0.129}_{-0.105}$ & 0.450$^{+0.100}_{-0.077}$ & 0.325$^{+0.028}_{-0.024}$ & 0.296$^{+0.031}_{-0.026}$\\[0.05in]
   $r_3$, \mum & 0.217$^{+0.013}_{-0.013}$ & 0.213$^{+0.013}_{-0.013}$ & 0.220$^{+0.012}_{-0.012}$ & 0.212$^{+0.013}_{-0.013}$\\[0.05in]
 $\tau_3$ & 0.330$^{+0.050}_{-0.046}$ & 0.293$^{+0.034}_{-0.032}$ & 0.351$^{+0.035}_{-0.033}$ & 0.289$^{+0.034}_{-0.032}$\\[0.05in]
$\tau_3/(\pi r_3^2 Q_\mathrm{ext})$, part./cm$^2$ &4.67$\times10^8$ & 4.57$\times10^8$ & 4.64$\times10^8$ & 4.56$\times10^8$\\[0.05in]
mass density, $\mu$g/cm$^2$ &  20.03 & 18.38 &  20.86 &  18.22\\[0.05in]
 $nh_3p_1$, bar & 1.520$^{+0.520}_{-0.340}$ & 1.460$^{+0.140}_{-0.120}$ & 1.620$^{+0.490}_{-0.340}$ & 1.500$^{+0.130}_{-0.120}$\\[0.05in]
$nh_3v_1$$\times 10^4$ & 0.426$^{+0.177}_{-0.128}$ & 0.436$^{+0.201}_{-0.141}$ & 0.363$^{+0.134}_{-0.100}$ & 0.357$^{+0.186}_{-0.124}$\\[0.05in]
$\chi^2$ &  96.04 &  87.03 &  90.49 &  89.28\\[0.05in]
$\chi^2/N$ &   1.17 &   1.06 &   1.10 &   1.09\\[0.05in]
\hline\\[-0.1in]
\end{tabular}\label{Tbl:lowphasevar}
\parbox{5.in}{Note: Latitudes in column headings are planetocentric. The \nht  mixing ratio was set to $nh_3v_0=4\times 10^{-4}$
for $p>p1$. Measured I/F values were multiplied by scale prior to finding best-fit parameter values.}
\end{table*}

\section{Results from fitting medium phase angle observations.}

As evident in Fig.\ \ref{Fig:medphase}, the medium phase angle
observations from 31 December 2000 and 1 January 2001 offer the
advantages of higher spatial resolution and two different viewing
geometries that provide additional constraints on the vertical
structure of clouds and hazes, as well as their scattering properties.
By accounting for the zonal wind-induced drift of features during the 38.85 hours
between the two observations, we are able to extract spectral samples
from the same atmospheric region, although whether this represents the
same cloud structure depends on whether it has evolved over that time
interval.  Thus, our constraint improvement is partly negated by the uncertain
degree of structure evolution that might have taken place.  There is
also the possibility that eddy motions have altered the position of
the feature from what we predicted using the assumption of a constant
zonal wind speed. In addition, there is some uncertainty in
the spectral sample due to the pixel quantization of the image,
as well as errors in navigation.  A partial assessment of whether our initial
assumption that the structure has not changed is true or not can be
made by comparing the fit quality obtained from fitting just one of
the spectra to that obtained from fitting both simultaneously. For this
purpose we would compare the reduced \chisq values for each alternative.
In most cases it appears that fit quality is not seriously degraded by
forcing the model to fit both observations simultaneously.

\begin{figure}\centering
\includegraphics[width=3.45in]{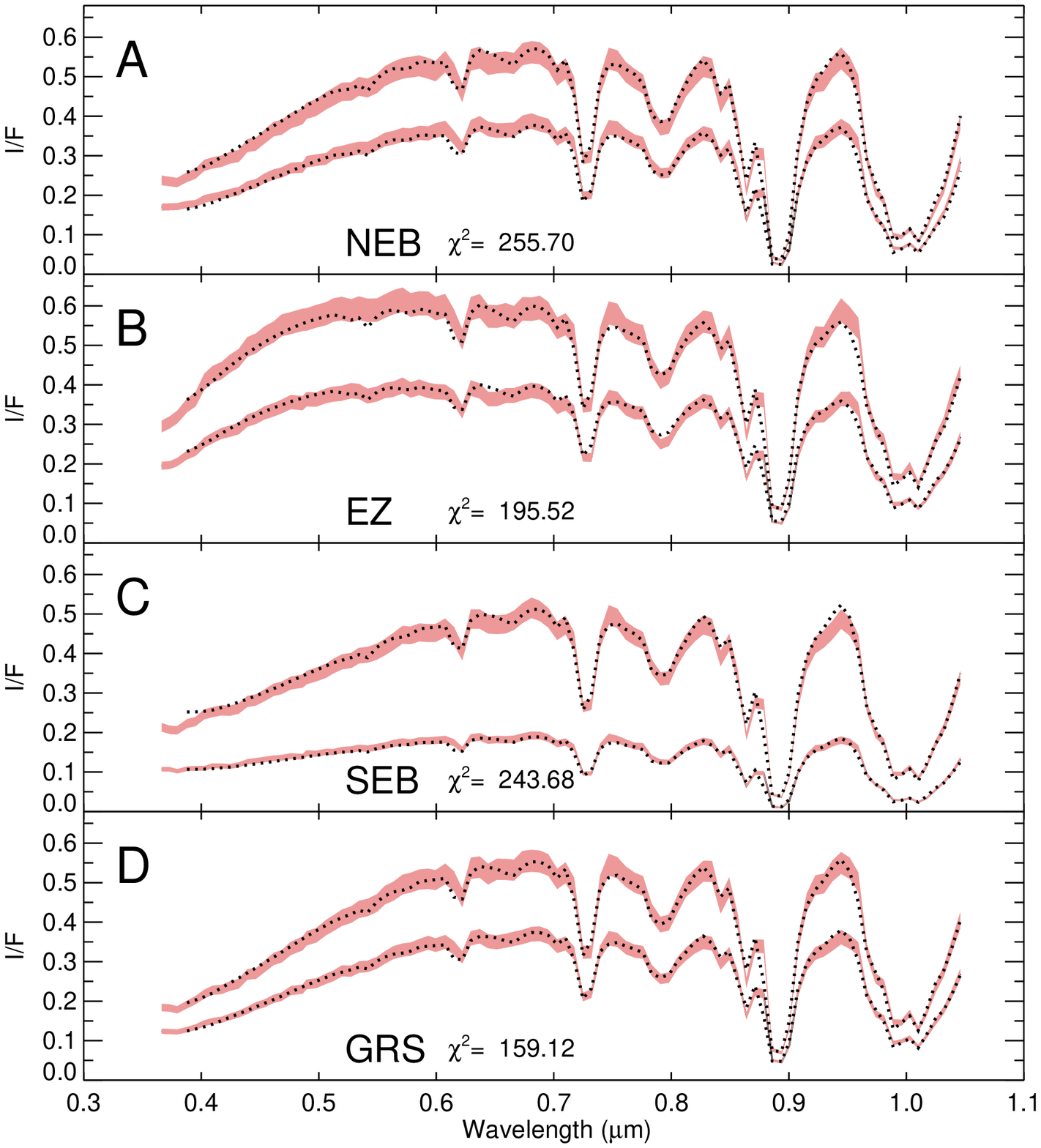}
\caption{\Cbm  model spectra for NEB (A), GRS (B), EZ (C), and SEB (D),
shown as dotted lines compared to the medium phase angle measured spectra, displayed as light red bands
bounding the assumed uncertainty range for combined model and measurement
errors. In each panel the lower spectrum is from 31 December 2000 and the upper from 2 January 2001.
For the selected features the larger incidence angle cosines in the January image produce brighter
spectra in spite of the smaller emission angle cosines.}
\label{Fig:fitmedspecs}
\end{figure}

The \cbm models that were fit to both spectral samples simultaneously
produced spectra that are shown in Fig.\ \ref{Fig:fitmedspecs} in
comparison with the measured spectra, using the same style as shown
for the low phase angle observations.  The parameter values derived
from these fits can be found in Table\ \ref{Tbl:medphasedual}. Similar
fits assuming $n_2$=1.85 yield results given in
Table\ \ref{Tbl:medphasedual1.85}. These are all very good fits, although
the NEB and SEB spectra are not as well fit as the GRS and EZ
spectra. This pattern was also seen in the low phase angle
observations, but differences in reduced \chisq values were much
smaller between different regions.  It is conceivable that the deeper
cloud structure of the NEB and SEB regions of Jupiter might mean that
the cloud top is actually deeper than the chromophore layer, and that
perhaps the \cbm model is not the most appropriate model for these
regions.  However, when we fit the spectra from just a single viewing
geometry (Table\ \ref{Tbl:medphase}), we don't see nearly as much
difference between the different regions. In fact, the reduced \chisq
values more closely resemble those obtained from the low phase angle
observations, as given in Table\ \ref{Tbl:lowphase}.  The difference
thus might be less due to problems with the model structure and more
due to sampling problems. In the next subsection we further consider what
constraints can be placed on the vertical structure of the chromophore
haze itself.

\subsection{Constraining the vertical location of the chromophores}

When we allowed the thin layer of chromophore particles to be elevated
above the main cloud layer, with the pressure allowed to be adjusted
to minimize \chisqx, the result for the low phase angle NEB
observation was that the pressure of the elevated haze was driven as
close to the top of the main cloud as the range boundaries allowed.
However, when we forced the chromophore to reside at 60 mbar instead
of putting it at the top of the main cloud layer (near 200 mb), we got
an excellent overall fit that was only slightly deviant at the
shortest wavelengths where Rayleigh scattering is attenuated by the
high chromophore layer. Moving the layer downward would provide a
useful I/F boost that would improve the fit at short wavelengths.

Using the more constrained set of
observations near the end of December, which provide two different
viewing geometries, we got a more definitive result.  When we put the
chromophore into a diffuse stratospheric haze layer extending from the
top of the main cloud upward, and then tried to find the optimum ratio
of the scale height of that haze to the pressure scale height, the
scale height dropped to the minimum value allowed.  The implication is
that the chromophore layer seems to fit best when it is very thin, and
right on top of the main cloud layer. 

Considering the GRS dual spectral observations at medium phase angle to be the most
reliably located with respect to the target feature in both cases, we first tried
to use those spectra to better constrain the  haze of chromophores above the GRS.  Using 
a detached haze of chromophores again, we tried to optimize the fit to the dual spectra
with an additional parameter, namely the pressure of the chromophore layer, in
this case not fixing it to the top of main cloud layer but allowing it to move
between 200 mbar and 40 mbar.  The result was that the haze pressure was again forced
to the maximum value allowed, which is so close to the top of the main cloud
that we cannot really distinguish it from being in contact with the main cloud.
In this case we did not provide for a stratospheric haze at 40 mbar that might
provide a positive scattering contribution. So, perhaps the haze was partly 
forced downward to allow a short wavelength boost of Rayleigh scattering.  However, when
we added such a haze, we got essentially the same result.  
This is similar to the results
of \cite{Baines2016Icarus}.  When they put the chromophore into a
stratospheric haze at 40 mbar, their fit quality was worse, but not
dramatically so (\chisqx /N increased from 0.92 to 1.2).

\begin{table*}\centering
\caption{Best-fit parameter values assuming $n_2$=1.40 for \cbm model fits simultaneously
  constrained by VIMS medium phase-angle observations from both
  viewing geometries shown in Fig.\ \ref{Fig:medphase}.}
\begin{tabular}{ r c c c c}
Parameter, unit &     GRS (20.4\deg S) &     SEB (14.7\deg S) &       EZ (0.7\deg N) &     NEB (13.3\deg N)\\[0.05in]
\hline\\[-0.06in]
               $\tau_1$ & 0.004$^{+0.004}_{-0.002}$ & 0.027$^{+0.003}_{-0.003}$ & 0.000$^{+0.000}_{-0.000}$ & 0.010$^{+0.004}_{-0.003}$\\[0.05in]
          $p_{2\mathrm{T}}$, bar & 0.205$^{+0.012}_{-0.012}$ & 0.489$^{+0.018}_{-0.018}$ & 0.060$^{+0.034}_{-0.011}$ & 0.381$^{+0.017}_{-0.017}$\\[0.05in]
           $p_2$, bar & 4.192$^{+0.630}_{-0.614}$ & 4.900$^{+0.703}_{-0.741}$ & 2.154$^{+0.151}_{-0.138}$ & 3.213$^{+0.243}_{-0.230}$\\[0.05in]
               $\tau_2$ & 29.56$^{+ 4.24}_{- 3.76}$ &25.187$^{+4.179}_{-3.631}$ &13.663$^{+1.348}_{-1.232}$ &16.061$^{+1.057}_{-0.995}$\\[0.05in]
$\tau_2/(p_2-p_{2\mathrm{T}})$, optical depth/bar &    7.41 &   5.71 &   6.52 &   5.67\\[0.05in]
          $r_2$, \mum & 1.148$^{+0.277}_{-0.245}$ & 0.836$^{+0.239}_{-0.193}$ & 0.586$^{+0.066}_{-0.059}$ & 1.438$^{+0.220}_{-0.213}$\\[0.05in]
          $r_3$, \mum & 0.149$^{+0.007}_{-0.007}$ & 0.286$^{+0.014}_{-0.015}$ & 0.117$^{+0.014}_{-0.012}$ & 0.151$^{+0.010}_{-0.009}$\\[0.05in]
               $\tau_3$ & 0.209$^{+0.021}_{-0.020}$ & 0.757$^{+0.058}_{-0.064}$ & 0.059$^{+0.012}_{-0.010}$ & 0.186$^{+0.023}_{-0.021}$\\[0.05in]
$\tau_3/(\pi r_3^2 Q_\mathrm{ext})$, part./cm$^2$ &1.79$\times 10^9$ & 3.18$\times10^8$ & 1.70$\times 10^9$ & 1.49$\times 10^9$\\[0.05in]
mass density, $\mu$g/cm$^2$ & 24.71 & 31.15 & 11.34 & 21.46\\[0.05in]
        $nh_3p_1$, bar & 1.950$^{+1.580}_{-0.840}$ & 2.020$^{+0.590}_{-0.460}$ & 1.280$^{+0.140}_{-0.120}$ & 2.180$^{+0.800}_{-0.610}$\\[0.05in]
$nh_3v_1$$\times 10^4$ & 0.988$^{+0.382}_{-0.302}$ & 0.449$^{+0.103}_{-0.085}$ & 0.703$^{+0.477}_{-0.308}$ & 0.494$^{+0.093}_{-0.080}$\\[0.05in]
$\chi^2$ & 159.12 & 243.68 & 195.52 & 255.70\\[0.05in]
$\chi^2/N$ &   0.97 &   1.49 &   1.19 &   1.56\\[0.05in]
\hline\\[-0.1in]
\end{tabular}\label{Tbl:medphasedual}
\parbox{5in}{NOTE: Latitudes in column headings are planetocentric. \nht mixing ratio $nh_3v_1$ and pressure $nh_3p_1$ were derived assuming $nh_3v_0=4\times 10^{-4}$.}
\end{table*}
\begin{table*}\centering
\caption{Best-fit parameter values assuming $n_2$=1.85 for \cbm model fits simultaneously
  constrained by VIMS medium phase-angle observations from both
  viewing geometries shown in Fig.\ \ref{Fig:medphase}.}
\begin{tabular}{ r c c c c}
Parameter, unit &     GRS (20.4\deg S) &     SEB (14.7\deg S) &       EZ (0.7\deg N) &     NEB (13.5\deg N)\\[0.05in]
\hline\\[-0.05in]
               $\tau_1$ & 0.003$^{+0.005}_{-0.002}$ & 0.026$^{+0.003}_{-0.003}$ & 0.005$^{+0.040}_{-0.004}$ & 0.005$^{+0.014}_{-0.004}$\\[0.05in]
          $p_{2\mathrm{T}}$, bar & 0.198$^{+0.012}_{-0.011}$ & 0.486$^{+0.018}_{-0.018}$ & 0.125$^{+0.017}_{-0.014}$ & 0.270$^{+0.013}_{-0.013}$\\[0.05in]
           $p_2$, bar & 3.887$^{+0.493}_{-0.471}$ & 4.687$^{+0.647}_{-0.663}$ & 2.826$^{+0.208}_{-0.194}$ & 2.907$^{+0.259}_{-0.240}$\\[0.05in]
               $\tau_2$ & 14.11$^{+ 2.07}_{- 1.82}$ &10.857$^{+2.001}_{-1.698}$ &13.890$^{+0.809}_{-0.766}$ & 8.190$^{+0.811}_{-0.739}$\\[0.05in]
$\tau_2/(p_2-p_{2\mathrm{T}})$, optical depth/bar &    3.83 &   2.58 &   5.14 &   3.11\\[0.05in]
          $r_2$, \mum & 0.416$^{+0.063}_{-0.052}$ & 0.329$^{+0.046}_{-0.037}$ & 0.988$^{+0.049}_{-0.047}$ & 0.346$^{+0.035}_{-0.030}$\\[0.05in]
          $r_3$, \mum & 0.140$^{+0.007}_{-0.007}$ & 0.286$^{+0.013}_{-0.014}$ & 0.126$^{+0.015}_{-0.013}$ & 0.278$^{+0.016}_{-0.017}$\\[0.05in]
               $\tau_3$ & 0.181$^{+0.016}_{-0.015}$ & 0.744$^{+0.056}_{-0.061}$ & 0.063$^{+0.015}_{-0.012}$ & 0.545$^{+0.053}_{-0.053}$\\[0.05in]
$\tau_3/(\pi r_3^2 Q_\mathrm{ext})$, part./cm$^2$ &2.13$\times 10^9$ & 3.12$\times10^8$ & 1.23$\times 10^9$ & 2.57$\times10^8$\\[0.05in]
mass density, $\mu$g/cm$^2$ & 24.274 & 30.603 & 10.292 & 23.168\\[0.05in]
        $nh_3p_1$, bar & 2.010$^{+2.020}_{-0.990}$ & 2.000$^{+0.630}_{-0.470}$ & 1.170$^{+2.740}_{-0.437}$ & 1.990$^{+0.940}_{-0.630}$\\[0.05in]
$nh_3v_1$$\times 10^4$ & 1.150$^{+0.600}_{-0.450}$ & 0.417$^{+0.105}_{-0.085}$ & 2.010$^{+1.360}_{-1.360}$ & 0.490$^{+0.162}_{-0.125}$\\[0.05in]
$\chi^2$ & 164.34 & 240.53 & 194.55 & 180.40\\[0.05in]
$\chi^2/N$ &   1.00 &   1.47 &   1.19 &   1.10\\[0.05in]
\hline\\[-0.1in]
\end{tabular}\label{Tbl:medphasedual1.85}
\parbox{5in}{NOTE: Latitudes in column headings are planetocentric. \nht mixing ratio $nh_3v_1$ and pressure $nh_3p_1$ were derived assuming $nh_3v_0=4\times 10^{-4}$.}
\end{table*}

\begin{table*}\centering
\caption{Best-fit parameter values assuming $n_2$=1.40  for \cbm model fits to just the VIMS medium
  phase-angle observations from the single observing geometry on 31 December 2000.}\vspace{0.01in}
\begin{tabular}{ r c c c c}
Parameter, unit &     GRS (20.4\deg S) &     SEB (14.7\deg S) &       EZ (0.7\deg N) &     NEB (13.5\deg N)\\[0.05in]
\hline\\[-0.05in]
               $\tau_1$ & 0.007$^{+0.008}_{-0.004}$ & 0.022$^{+0.003}_{-0.003}$ & 0.000$^{+0.000}_{-0.000}$ & 0.021$^{+0.010}_{-0.007}$\\[0.05in]
          $p_{2\mathrm{T}}$, bar & 0.204$^{+0.017}_{-0.015}$ & 0.443$^{+0.021}_{-0.021}$ & 0.072$^{+0.057}_{-0.016}$ & 0.383$^{+0.028}_{-0.027}$\\[0.05in]
           $p_2$, bar & 3.996$^{+0.964}_{-0.901}$ & 4.534$^{+0.901}_{-0.912}$ & 2.656$^{+0.338}_{-0.299}$ & 3.486$^{+0.439}_{-0.409}$\\[0.05in]
               $\tau_2$ & 32.74$^{+11.68}_{- 8.89}$ &22.311$^{+4.504}_{-3.797}$ &19.460$^{+2.608}_{-2.318}$ &18.837$^{+2.014}_{-1.831}$\\[0.05in]
$\tau_2/(p_2-p_{2\mathrm{T}})$, optical depth/bar &    8.63 &   5.45 &   7.53 &   6.07\\[0.05in]
          $r_2$, \mum & 0.655$^{+0.368}_{-0.234}$ & 2.365$^{+0.066}_{-0.071}$ & 0.764$^{+0.212}_{-0.169}$ & 1.201$^{+0.328}_{-0.290}$\\[0.05in]
          $r_3$, \mum & 0.137$^{+0.017}_{-0.015}$ & 0.217$^{+0.030}_{-0.029}$ & 0.117$^{+0.036}_{-0.025}$ & 0.140$^{+0.017}_{-0.015}$\\[0.05in]
               $\tau_3$ & 0.192$^{+0.034}_{-0.030}$ & 0.492$^{+0.101}_{-0.097}$ & 0.050$^{+0.028}_{-0.018}$ & 0.159$^{+0.032}_{-0.027}$\\[0.05in]
$\tau_3/(\pi r_3^2 Q_\mathrm{ext})$, part./cm$^2$ &2.51$\times 10^9$ & 7.02$\times10^8$ & 1.42$\times 10^9$ & 1.86$\times 10^9$\\[0.05in]
mass density, $\mu$g/cm$^2$ &  26.83 &  29.94 &  9.51 & 21.27\\[0.05in]
       $nh_3p_1$, bar & 2.050$^{+0.590}_{-0.460}$ & 1.810$^{+0.280}_{-0.240}$ & 1.130$^{+0.940}_{-0.313}$ & 1.950$^{+0.820}_{-0.530}$\\[0.05in]
$nh_3v_1$$\times 10^4$ & 0.868$^{+0.542}_{-0.371}$ & 0.400$^{+0.153}_{-0.112}$ & 0.703$^{+0.847}_{-0.427}$ & 0.388$^{+0.117}_{-0.092}$\\[0.05in]
$\chi^2$ &  82.07 &  97.94 &  78.76 & 103.60\\[0.05in]
$\chi^2/N$ &   1.00 &   1.19 &   0.96 &   1.26\\[0.05in]
\hline\\[-0.1in]
\end{tabular}\label{Tbl:medphase}
\parbox{5in}{NOTE: Latitudes in column headings are planetocentric. Locations of spectral samples are shown in Fig.\ \ref{Fig:medphase}A. \nht mixing ratio $nh_3v_1$ and pressure $nh_3p_1$ were derived assuming $nh_3v_0=4\times 10^{-4}$ and $nh_3rh=1.0$.}
\end{table*}

\begin{figure*}\centering
\includegraphics[width=5.2in]{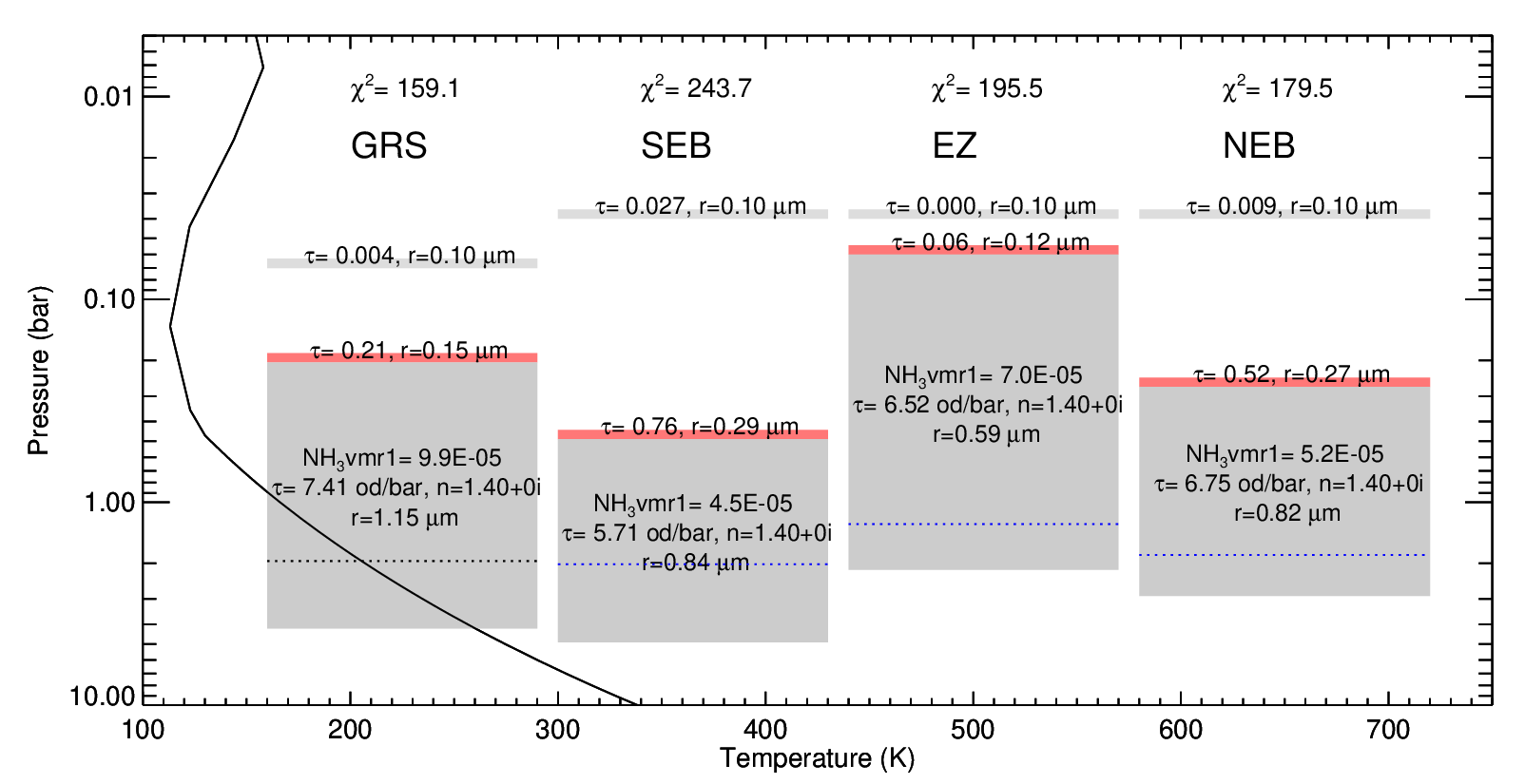}\vspace{-0.1in}
\includegraphics[width=5.2in]{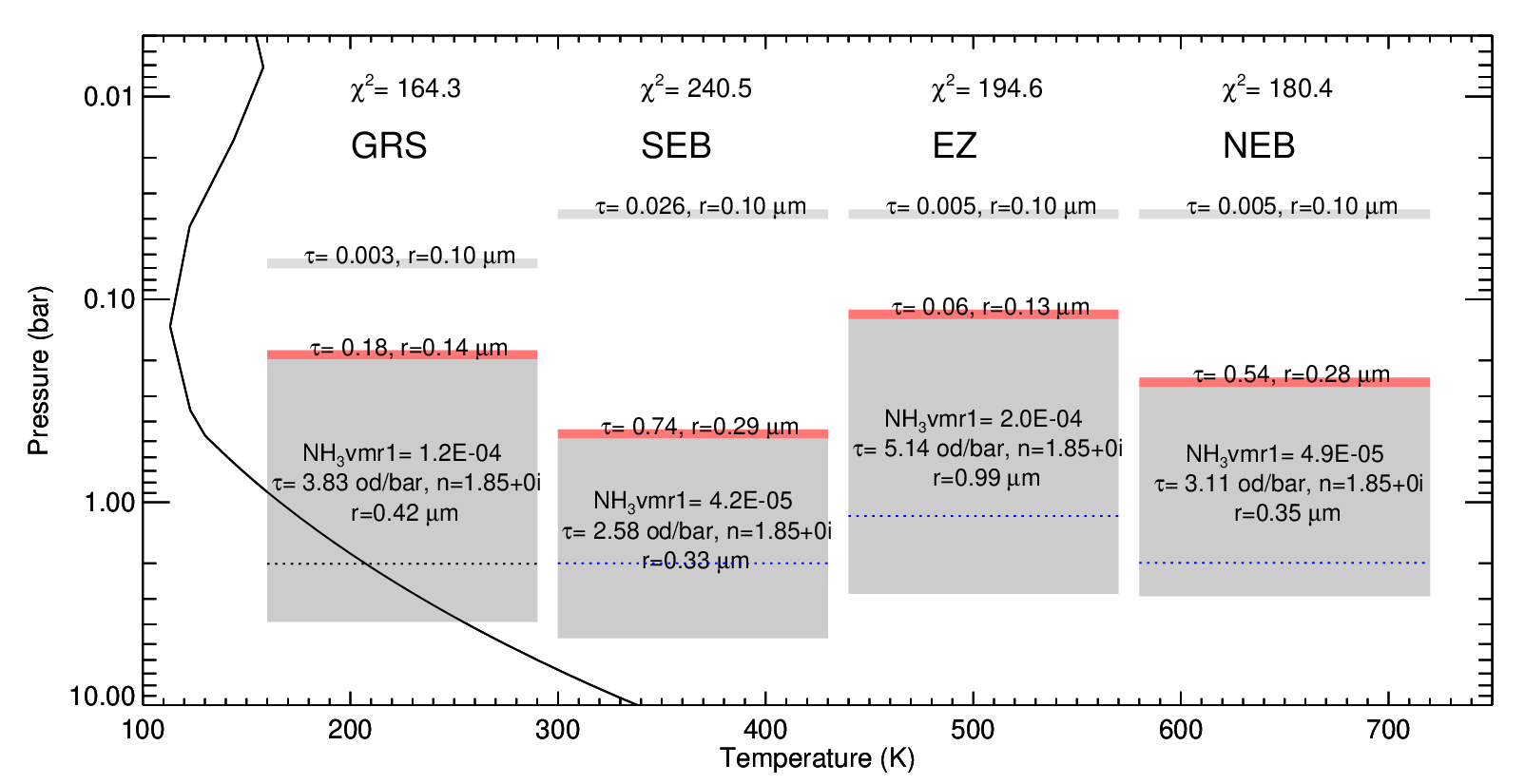}
\caption{\Cbm cloud model fits to medium phase-angle VIMS spectral
  samples from NEB, GRS, EZ, and SEB regions on Jupiter, using
  constraints from both viewing geometries.  The top panel is for
  $n_2=1.4+0i$ and the bottom for $n_2=1.85+0i$. The higher index has
  the advantage of keeping the top of the main cloud in the equatorial
  zone below the the temperature minimum.  The blue dotted line
  indicates the pressure below which the \nht vmr is assumed to be
  $4\times10^{-4}$.  Spectral fits are shown in
  Fig.\ \ref{Fig:fitspecs}. Adjusted parameter values and
  uncertainties are given in Table\ \ref{Tbl:medphasedual} ($n_2$=1.4)
  and Table \ref{Tbl:medphasedual1.85} ($n_2$=1.85).}
\label{Fig:medcartoon}
\end{figure*}

\subsection{Ammonia fit results}

The most consistent result from both low and high phase angle VIMS
observations is that the \nht VMR immediately below the condensation level is
significantly higher in the GRS and EZ by about a factor of two than
it is in the SEB and NEB.  The latter values range from 4.2$\times
10^{-5}$ to 4.9$\times 10^{-5}$, with an average of 4.5$\times
10^{-5}$, which corresponds to an ammonia condensation level near 585
mbar, which is well below the cloud tops for all regions,
and also well above the cloud bottoms. So the part of the NEB cloud
deeper than 585 mbar is most likely composed of \nhfshx, while the
upper part of the cloud is likely a mix of \nht and \nhfshx.  Whether
such a construct would fit the observations much better remains to be
seen, but it is clear that our conclusions about the suitability of
the \cite{Carlson2016} chromophore as a coloring agent on Jupiter
would not be affected.  There is less consistency and much more
uncertainty in the depth at which the upper mixing ratio transitions
to the deeper mixing ratio. Pressures from 1.43 bars to 1.65 bars are
inferred from the low phase angle observations, but the uncertainties
are very large, about 0.3 bars in the direction of lower pressures and
0.4 to 0.5 bars in the direction of increasing pressure. The
transition pressures for the high phase angle dual fits are larger,
averaging 2.1 bars with somewhat larger uncertainties in both
directions.

The ammonia mixing ratios immediately below the condensation level in the GRS average about
9$\times 10^{-5}$ with uncertainties of about 0.3-0.4$\times 10^{-5}$.  This corresponds
to a condensation pressure near 634 mbar, also well below the cloud top
and well above the cloud bottom.  Thus the GRS is likely of mixed composition as well,
though a bigger fraction of the cloud is at higher altitudes relative to the SEB or NEB.
The EZ has a sub-condensation mixing ratio comparable to or lower than that of the GRS,
but transitions to the deep mixing ratio at a lower pressure.

\section{Speculation on Physical Mechanisms}

Production of the \cite{Carlson2016} chromophore depends on UV flux,
ammonia, and acetylene.  Because ammonia falls off with altitude and
UV flux increases with altitude it is expected that production of
photolyzed ammonia would occur somewhat above the cloud tops and
would be widespread over Jupiter. The availability of
acetylene seems to be a controlling factor in the production rate and
its flux is quite weak on average according to current photochemical models
\citep{Moses2010FaDi}. \cite{Baines2016Icarus} argued that lightning
might raise the local acetylene mixing ratio sufficiently
to produce the chromophore amounts needed at the top of the
GRS.  However, that argument would not work to explain the
widespread distribution of the chromophore, as seems to be indicated
by the close spectral matches we found over a number of different regions.
An alternative suggested by \cite{Carlson2016} is that there may be an
important photochemical role of ice grains and polyacetylene aerosols.
If ammonia ice grains are important, that might explain why the
chromophore seems to be located at the cloud tops (it might be
produced there).

The variations we have seen in aerosol properties among different
features, which are also at different latitudes, might have to do with
stratospheric and/or upper tropospheric dynamics or differences in
eddy mixing in the vicinity of the cloud tops. The biggest difference
we have seen overall in chromophore mass loading is at the top of the
Equatorial Zone, where we find about half the amounts seen in other
regions.  This might be evidence against the ice grain mechanism,
considering that the cloud top is significantly higher and would be
exposed to more UV flux, suggesting more production of chromophores
than average, not less.  This might be evidence instead for the gas
phase production, because the ammonia mixing ratio above this high
cloud feature would be less than average (assuming that ammonia is
falling off with altitude at the same rate at all locations).
However, both mechanisms depend on other factors such as eddy
mixing differences that might work in the opposite direction.  It
remains to be determined whether there is a plausible mechanism to
produce the chromophore amounts that are needed to match VIMS spectra
and to explain the variations among different features.  Photochemical
models, microphysical models of cloud particle evolution, and
dynamical models all seem needed to reach an understanding.

\section{Summary and Conclusions}

We used Jupiter's 0.35-1.1 \mum spectrum, as measured by the Cassini/VIMS instrument
near the end of 2000, to constrain cloud structures for the GRS, the equatorial zone,
and north and south equatorial belts. We used a simple model structure in which
the main cloud was composed of conservative particles and covered by a thin layer
of particles made of the chromophore of \cite{Carlson2016}. Our main conclusions
from this investigation are as follows.

\begin{enumerate}

\item The substance described by \cite{Carlson2016} appears to be a
  universal chromophore for Jupiter's clouds. Among the four cloud
  regions we studied with low phase angle observations, all have a
  reddish color that can be reproduced with the same kind of \cbm
  model that \cite{Baines2016Icarus} used to model the GRS spectrum, i.e. a
  physically and optically thin layer immediately above the main cloud
  layer, with modest variations in particle size (from 0.13 \mum for the
  GRS to 0.22 \mum for the NEB) and, in all but one case, modest
  variations in 1-\mum optical depth (from 0.13 for the GRS to 0.33 for the
  NEB).  For the medium phase angle observations, the range is similar
except that the SEB spectrum in that data set led to the largest particle
size (0.286 \mumx) and the largest optical depth (0.76).

\item The efficacy of the \cite{Carlson2016} chromophore in
  reproducing Jovian spectral colors is robust, even for 12\% changes
  in VIMS calibration and large uncertainties in the refractive index
  of the main cloud layer due to uncertain fractions of \nhfsh and
  \nht in its cloud particles.

\item Medium phase angle VIMS observations, which provided two different view
and illumination angles of the same cloud regions, provide additional
constraints on vertical structure, but add uncertainty due to the
38.5-hour time difference between observations. Although these dual-angle
fits produced about the same fit quality and similar
properties of the chromophore layer, they yielded much larger particles
for the main cloud layer (0.6 to 1.4 \mumx, compared to 
0.3-0.6 \mum obtained from the low phase angle observations).

\item According to the low phase angle observations, the vertically
  integrated masses of chromophore material above the GRS, SEB, EZ,
  and NEB are remarkably similar, ranging only from 13 to 20
  $\mu$g/cm$^2$, and only 18 to 20 $\mu$g/cm$^2$
  if the EZ value is excluded.  The range was somewhat larger
for the medium phase angle results, ranging from 11 $\mu$g/cm$^2$
for the EZ to 21 -- 31 $\mu$g/cm$^2$ for the other regions.

\item We also found a depression of the ammonia volume mixing
ratio in the two belt regions, which averaged 0.4-0.5$\times 10^{-4}$
immediately below the ammonia condensation level, while the other
regions averaged twice that value.

\end{enumerate}

\section*{Acknowledgments}

We thank Gordon Bjoraker and an anonymous reviewer for constructive reviews.
LAS and PMF acknowledge support by NASA grant NNX14AH40G from NASA's Planetary Atmospheres Program.



\end{document}